\documentclass[aip,manuscript]{revtex4-1}
\usepackage{graphicx}
\usepackage{epsfig}
\usepackage{subfigure} 
\usepackage{amssymb,amsmath}
\usepackage{overpic}
\usepackage{epstopdf}

\begin{document}

\title{Intradomain phase transitions in flexible block copolymers with self-aligning segments} 



\author{Christopher J. Burke}
\email[]{cburke@mail.pse.umass.edu}
\affiliation{Department of Polymer Science, University of Massachusetts, Amherst, Massachusetts 01003, USA}

\author{Gregory M. Grason}
\email[]{grason@mail.pse.umass.edu}
\affiliation{Department of Polymer Science, University of Massachusetts, Amherst, Massachusetts 01003, USA}



\begin{abstract}
  We study a model of flexible block copolymers (BCPs) in which there is an enlthalpic preference for orientational order, or local alignment, among like-block segments. We describe a generalization of the self-consistent field theory (SCFT) of flexible BCPs to include inter-segment orientational interactions via a Landau-DeGennes free energy associated with a polar or nematic order parameter for segments of one component of a diblock copolymer. We study the equilibrium states of this model numerically, using a pseudo-spectral approach to solve for chain conformation statistics in the presence of a self-consistent torque generated by inter-segment alignment forces. Applying this theory to the structure of lamellar domains composed of symmetric diblocks possessing a single block of ``self-aligning", polar segments, we show the emergence of spatially complex segment order parameters (segment director fields) within a given lamellar domain.  Because BCP phase separation gives rise to spatially inhomogeneous orientation order of segments even in the absence of explicit intra-segment aligning forces, the director fields of BCPs, as well as thermodynamics of lamellar domain formation, exhibit a highly non-linear dependence on both the inter-block segregation ($\chi N$) and the enthalpy of alignment ($\varepsilon$).  Specifically, we predict the stability of new phases of lamellar order in which distinct regions of alignment {\it coexist} within the single mesodomain, and which spontaneously break the symmetries of the lamella (or smectic) pattern of composition in the melt via in-plane tilt of the director in the centers of the like-composition domains.  We show further that, in analogy to a Freedericksz transition confined nematics, that the elastic costs to reorient segments within the domain, as described by Frank elasticity of the director, increase the threshold value $\varepsilon$ needed to induce this {\it intra-domain phase transition}.
\end{abstract}


\maketitle 

\section{Introduction}

Block copolymers (BCPs) are polymer molecules consisting of blocks of different monomer species, whose generic immiscibility gives rise to the tendency for inter-block phase separation in melts, and ultimately, the formation of a rich spectrum of periodically-ordered mesodomain structures in equilibrium\cite{Bates1990}.  The morphology of these mesodomains, as well as the symmetries of their inter-domain arrangement, depends on the chemical properties of the monomers and the chain structure and topology\cite{Grason2006,Matsen2012}. Self-consistent (mean) field theory (SCFT) has been widely used to model this phase separation process. The most basic SCFT model accounts for monomer immiscibility via a Flory-Huggins interaction which assumes isotropic pair interactions, and accounts for the chain conformation statistics assuming fully-flexible (Gaussian) chain, in which unlike segments repel through short-ranged, isotropic interactions (e.g. Flory-Huggins model) \cite{Leibler1980,Matsen2012,Helfand1975}. The flexible-chain, Flory-Huggins (FFH) model is successful in modeling the size and structure of mesodomain formation, and in large part, the phase diagram of melts; however, quantitative discrepancies between phase boundaries and experimental systems are well known\cite{Matsen2002}. For example, polyisoprene-polystyrene diblocks exhibit a phase diagram which is asymmetric with respect to the monomer volume fraction, which undergoes an order-to-disorder transition at higher segregation strengths than predicted, and which has a non-convex order-to-disorder phase boundary\cite{Khandpur1995}.  Such discrepancies have motivated several generalizations of this standard model to incorporate more chemically relevant details or experimentally realistic conditions, including the effect of strong fluctuations at weak segregation\cite{Fredrickson1987}, the effects of segmental (conformational) asymmetry\cite{Matsen1994,Milner1994}, and the effects of polydispersity in melts\cite{Burger1990}.  In this article, we consider a flexible-chain model of BCPs generalized to include non-isotropic segmental interactions.

Broadly speaking, we can divide orientational interaction models into two categories, intra-chain and inter-chain segment interactions.  The former case is best known in the context of persistent chain models (e.g. semi-flexible or rod-like chains) where backbone stiffness favors alignment along a given polymer block.  Inter-chain effects, on the other hand, model the entropic (e.g. Onsager model) and enthalpic (e.g. Maier-Saupe model) dependence on the relative orientation on anisotropic elements (chain segments) on the local packing in a concentrated, multichain system.  While liquid crystallinity in BCPs has received considerable theoretical attention to date~\cite{Chen2016}, to an overwhelming extent, previous studies consider the combined effects of intra- and inter-chain orientational simultaneously, with considerably more attention going to cases rigid chains, where persistence lengths exceed block lengths.  Rod-coil BCP models assume one block is flexible while one is fully rigid, and the rigid rods exhibit liquid-crystalline orientational interactions\cite{Pryamitsyn2004,Matsen1998}.  Other BCP models consider semiflexible chains in which the persistence length is most often taken to be comparable to the chain length\cite{Matsen1996,Jiang2011,Jiang2013,Chen2016}. Maier-Saupe interactions have been used to model orientational interactions in semi-flexible and rigid BCPs\cite{Singh1994,Netz1996,Greco2016,Jiang2016} .  

While it is perhaps most natural to consider intra- and inter-chain orientation as intrinsically linked through the specific backbone structure, liquid crystalline ordering of segments in BCPs does not require the existence of intra-chain stiffness.  This was considered explicitly in SCFT models by Netz and Schick\cite{Netz1996} which considered weakly-persistent limits of semiflexible chain BCPs, where constituent blocks are multiple (i.e. $\gtrsim 1$) persistent lengths long, and nevertheless, predicted the existence of multiple symmetry-breaking transitions in meso-ordered BCPs driven by orientational (inter-chain) segmental interactions.  

In this paper, we consider the influence of intra-chain orientational interactions on BCP assembly, in the specific limit of {\it fully-flexible chains}, where the contour length of each block is long in comparison to their persistent lengths, or equivalently, is composed of a large number $N\gg1$ of Kuhn segments.  Although it can expected on general grounds that the strength of inter-segment orientational interactions between segments are relatively weaker as the backbone persistence, and hence the segment aspect ratio, is reduced, it is also true that BCP assemblies generically possess orientational segment order {\it even in the absence of any explicit inter- or intra-chain aligning tendencies}.  A recent study of the FFH model of BCP melts, using both SCFT and MD simulations of the Kremer-Grest model, shows that microphase separation into unlike polymer domains, in combination with chain connectivity, are sufficient to establish liquid crystalline segment order at the intra-domain scale\cite{Prasad2017}.  The segmental alignment induced by mesodomain formation in flexible BCP melts is shown to be both highly heterogenous, varying in direction and magnitude at the intra-domain scale, and non-linearly dependent on degree of segregation and domain morphology (e.g, spheres, cylinders, networks, lamella).  

The generic omnipresence of some degree of weakly-liquid crystalline segment order in every microphase-separated state raises a basic question about the susceptibility of these intra-domain textures to anisotropy in inter-segment forces.  To address this, we construct what is arguable the simplest generalization of the FFH model of BCP to include anisotropy in the local interactions between segments.  Specifically, we study the SCFT of a flexible (freely-jointed or Gaussian) chain model of diblock copolymers whose segments possess a preference for local {\it polar} or {\it nematic} alignment.  With this theory we consider how a drive for local alignment of segments competes with the intrinsic tendencies for spatial non-uniform segment alignment which arise due to composition gradients and chain connectivity alone\cite{Carton1990,Szleifer1989,Morse1994}.  

\begin{figure}
\includegraphics[]{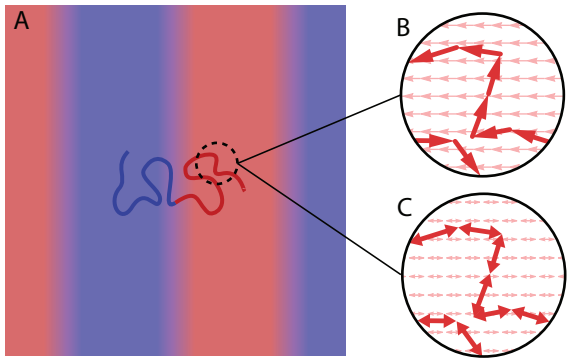}
\caption{Diblock copolymers phase separated into a lamellar mesophase. A) One polymer chain is shown, with the shaded background representing the average density of A (red) and B (blue) monomers. B-C) The chains are made up of many discrete segments whose size is smaller than the phase-separation lengthscale. The segments can be considered to have polar symmetry (B), pointing from the A-end to B-end of the chain, or they can have head-tail nematic symmetry (C). Chain segments can exhibit orientational interactions, which are accounted for via mean self-consistent torque fields which can be polar (B) or nematic in nature (C).}
  \label{fig:schematic}
\end{figure}

The interplay between periodic compositional order and orientational order in flexible-chain BCP represents a departure from the both previous models of liquid crystalline BCPs\cite{Pryamitsyn2004,Matsen1996,Netz1996,Li2016} as well as the ``canonical" picture of the coupling between nematic order and positionally ordered states of small-molecule liquid crystals~\cite{deGennes}.  In these cases, the size scale of the rigid blocks, or mesogenic units, is comparable to the domain size $D$ (say, the layer spacing of a smectic phase), and hence, the director field is uniform throughout a given domain (except at the core of topologically required defects) and strongly coupled to composition gradients.  In contrast, as shown in fig. \ref{fig:schematic}, in flexible-chain BCP the scale of segments, the Kuhn length $a$ is well separated from the domain size, $D\gg a$.  Due to increasing importance of thermal fluctuations for stretched chains as $a$ decreases, the magnitude of orientational order of segments decreases with increasing $N$ (with $\chi N$ held fixed), indicating a relatively weak coupling between the segment direction and composition gradients in the $N \gg 1$ regime and the orientational order parameter varies in magnitude and direction throughout the domain\cite{Prasad2017}.  This flexible-chain limit, as we describe below, gives rise to new regimes of spatially heterogeneous sensitivity of inter-segment aligning forces, and correspondingly new states of spatially modulated orientational at the sub-domain scale.

In this paper, we focus on the effects of locally aligning segment interactions in lamellar mesophases of flexible BCP. We begin by presenting an extension to the standard flexible-Flory-Huggins SCFT which includes local orientational interactions via a Landau model of the orientational enthalpy density in a diblock with one block of ``self-aligning" segments. We focus in detail on the case of interactions decribed by a polar order parameter, and discuss some aspects of the theory for nematic interactions.  We describe a numerical implementation of this model which employs a pseudo-spectral method, focusing on the challenges introduced by polar interactions, which introduce {\it self-consistent torque fields} that modify the random-walk statistics of polymeric blocks.

Based on the results of this new algorithm, we present phase diagrams for equal composition lamella with varying segregation strength ($\chi N$) and enthalpy of alignment of chain segments ($\varepsilon$). We find that increasing $\epsilon$ above a critical value leads to a symmetry breaking transition, where disordered cores of the standard smectic-A (SmA) lamellar morphology becomes unstable to spontaneous alignment parallel to the layer, leading to a two varients of a smectic-C like (SmC) state.  The tilt of the order parameter in these SmC phases occurs only within discrete sub-domains; we show that these subdomains constitute a distinct orientation-dominated phase which coexists with a segmental order induced by composition gradients and chain connectivity alone. We also find that for large enthalpy of alignment, the lamellar spacing exhibits a non-monotonic dependence on $\chi N$, and it grows anomalously as lamella approach the homogeneous phase. Finally, we discuss the role of free energy terms which penalize gradients in the segment orientation in the form of Frank elasticity for the director. Not unlike the Freedericksz transition in liquid crystal layers, increased Frank elasticity leads to a suppression of the spontaneous reorientation of segments in the SmC state, and consequent increase in the critical $\varepsilon$ that separates the SmA from the SmC phase.

\section{Theory}

\subsection{Orientational Self-Consistent Field Theory for Flexible BCPs} \label{sec:scfTheory}

Consider an AB diblock copolymer melt, consisting of $n$ chains occupying a volume $V$, each chain with $N$ segments of Kuhn length $a$. For concreteness, here, we consider a freely-jointed chain model, in which rigid chain segments connected end-to-end rotate freely relative to one another, although it is straight forward to generalize the derivation below to other flexible chain models, such as the Gaussian thread model~\cite{Matsen2012}, provided a suitable generalization of the local segment director~\cite{Zhao2012, Prasad2017}.  The monomer fraction of species A is $f$, and in the present study we consider chains in which only the like-block interactions between A segments are anisotropic. The immiscibility of monomers A and B is quantified by a Flory-Huggins parameter $\chi$, giving interaction enthalpy per chain (in units of $k_\mathrm{B}T$),
\begin{equation}
 F_{\mathrm{FH}}=\frac{N}{V}\int\mathrm{d}^3\mathbf{x}~\chi\phi_\mathrm{A}(\mathbf{x})\phi_\mathrm{B}(\mathbf{x})
\label{eqn:fScalar}
\end{equation}
where $\phi_{\mathrm{A}}(\mathbf{x})$ and $\phi_{\mathrm{A}}(\mathbf{x})$ are the local volume fractions of monomers A and B, respectively, acting as scalar order parameter fields.

In BCPs, due to their intrinsic ``head-tail" asymmetry, segment orientation can in general be described by two types of order parameters:  a vector order parameter $\mathbf{t}(\mathbf{x})$ to describe interactions with polar symmetry, or a tensor order parameter $Q_{ij}(\mathbf{x})$ to describe interactions with nematic symmetry. These order parameters describe the local average orientation of monomer A/B segments, defined as follows:
\begin{equation}
   \mathbf{t}_{A/B}(\mathbf{x})=\rho_0^{-1}\sum_{\alpha\in\mathrm{A/B}}\hat{\mathbf{r}}_\alpha\delta(\mathbf{x}-\mathbf{x}_\alpha)
   \label{eqn:tDef}
\end{equation}
\begin{equation}
   Q^{A/B}_{ij}(\mathbf{x})=\rho_0^{-1}\sum_{\alpha\in\mathrm{A/B}}\left((\hat{\mathbf{r}}_\alpha)_i(\hat{\mathbf{r}}_\alpha)_j-\frac{\delta_{ij}}{3}\right)\delta(\mathbf{x}-\mathbf{x}_\alpha),
   \label{eqn:QDef}
\end{equation}
where $\rho_0=nN/V$ is the segment density and $\hat{\mathbf{r}}_\alpha$ is the unit tangent vector of segment $\alpha$ along the chain from the A terminus to the AB junction, and the sums are taken over all A segments.  In this paper, because we focus on anisotropic interactions in the A-block, we need only consider the order parameter of this block, and hereafter, use the notation $\mathbf{t}(\mathbf{x}) \equiv \mathbf{t}_{\mathrm{A}}(\mathbf{x})$  and ${\bf Q}(\mathbf{x}) \equiv {\bf Q}_\mathrm{A}(\mathbf{x})$.  

We first describe the theory for BCPs with polar interactions, in which orientational interactions that favor local alignment are introduced via a Landau free energy associated with $\mathbf{t}(\mathbf{x})$. The orientational interaction enthalpy per chain is
\begin{equation}
   F_{\mathrm{local}}[{\bf t}({\bf x}) ] =\frac{N}{V}\int\mathrm{d}^3\mathbf{x}~\left(\frac{A}{2}|\mathbf{t}|^2+\frac{C}{4}|\mathbf{t}|^4\right).
   \label{eqn:polarEnthalpy}
\end{equation}
For $A\geq 0$, this enthalpy is minimized by $\mathbf{t}=0$, but for $A<0$, it is minimized by $|\mathbf{t}|=\sqrt{-A/C}$ in the absence of segment connectivity and at zero temperature. For negative values of $A$, the free energy can be parametrized in terms of the \emph{degree of preferred order} $t_0=\sqrt{-A/C}$, i.e. the value of $|\mathbf{t}|$ which minimizes the orientational enthalpy, and the \emph{enthalpy of alignment} $\epsilon=-A^2/4C$, i.e. the difference in the enthalpy per monomer between a melt in a disordered ($|\mathbf{t}|=0$) and a fully ordered ($|\mathbf{t}|=t_0$) state. This approach is similar to the Maier-Saupe model of orientation dependent enthalpy~\cite{deGennes}, with the difference here arising from the  polar rather nematic symmetry of interactions as well as the higher-terms ($|t|^4$) which describes an enthalpically preferred alignment of variable magnitude.

Beyond this preference for local alignment among like segments, the orientational inter-segment forces will also be described by elastic costs for distortions in segment director, as in previous work\cite{Zhao2012}. Including only gradient terms up to second order in ${\bf t}$~\footnote{Note that the total derivative term proportional to $\int\mathrm{d}^3\mathbf{x} ~(\nabla \cdot {\bf t})$ is also allowed by symmetry, but this vanishes for all periodic solutions of ${\bf t}({\bf x})$.}, this is described by the Frank elastic free energy,
\begin{equation}
   F_{\mathrm{grad}} [{\bf t}({\bf x}) ] =\frac{N}{2V}\int\mathrm{d}^3\mathbf{x}\left[K_1(\nabla\cdot\mathbf{t})^2+K_2(\nabla\times\mathbf{t})^2+2K_2q_0\mathbf{t}\cdot(\nabla\times\mathbf{t})\right]
   \label{eqn:gradientEnthalpy}
\end{equation}
where $K_1$ (splay) and $K_2$ (bend and twist) are elastic constants and $q_0$ is a chirality parameter, inversely related to the preferred chiral pitch $p=2\pi/q_0$, allowed for segments which lack an inversion symmetry, and used previously to model chirality transfer in block copolymers~\cite{Zhao2013}.

We incorporate these orientational interactions into the standard self-consistent field theory of BCP melts \cite{Matsen2002} following the same procedure as in ref. \cite{Zhao2012}. Segment-segment interactions are replaced with scalar self-consistent fields $w_\mathrm{A}$ and $w_\mathrm{B}$ which account for scalar density-dependent interactions, and vector mean field $\mathbf{W}$, a self-consistent torque, which represents the mean-field oreintational interactions among A segments. The full free energy per chain of a BCP melt with polar interactions is 
\begin{equation}
F=F_{\mathrm{FH}}+F_{\mathrm{p}}-TS.
\label{eqn:fullPolarFreeEnergy}
\end{equation}
Where $F_{\mathrm{p}}[{\bf t}({\bf x}) ] =F_{\mathrm{local}}[{\bf t}({\bf x}) ] +F_{\mathrm{grad}}[{\bf t}({\bf x}) ]$ is the orientational free energy of A-block segments. The entropy $S$ is given by 
\begin{equation}
S=\ln{\mathcal{Q}}+\frac{N}{V}\int\mathrm{d}^3\mathbf{x}\left(w_\mathrm{A}\phi_\mathrm{A}+w_\mathrm{B}\phi_\mathrm{B}+\mathbf{W}\cdot\mathbf{t}\right)
\label{eqn:polarEntropy}
\end{equation}
where $\mathcal{Q}$ is the chain partition function. 

Considering the limit of $N\gg 1$, where flexible BCP are well modeled by Gaussian chain conformations, chain statistics are solved via a diffusion equation for a partial partition function $q(\mathbf{x},s)$, which gives the likelihood that a chain segment at arclength $s$ along the chain will be at position $\mathbf{x}$ in space, under the influence of these mean fields, given an even distribution at the $s=0$ end of the chain:
\begin{equation}
\frac{\partial q}{\partial s}=\begin{cases}
    \frac{N}{6}\left[a\nabla+\mathbf{W}(\mathbf{x})\right]^2q-Nw_\mathrm{A}(\mathbf{x})q & \text{if $s<f$}\\
    \frac{Na^2}{6}\nabla^2q-Nw_\mathrm{B}(\mathbf{x})q & \text{if $s>f$}
  \end{cases}
\label{eqn:diffusionEqn1}
\end{equation}
under the boundary condition $q(\mathbf{x},0)=1$. A conjugate partial partition function $q^\dagger$ describes chain diffusion from the $s=1$ end:
\begin{equation}
-\frac{\partial q^\dagger}{\partial s}=\begin{cases}
    \frac{N}{6}\left[a\nabla-\mathbf{W}(\mathbf{x})\right]^2q^\dagger-Nw_\mathrm{A}(\mathbf{x})q^\dagger & \text{if $s<f$}\\
    \frac{Na^2}{6}\nabla^2q^\dagger-Nw_\mathrm{B}(\mathbf{x})q^\dagger & \text{if $s>f$}
  \end{cases}
\label{eqn:diffusionEqn2}
\end{equation}
under the boundary condition $q(\mathbf{x},1)=1$. Appendix~\ref{appendix:chainStatistics} presents a derivation of these equations, as well as the modified diffusion equations for the nematic interactions model.  Notably, the presence of the self-consistent torques modifies the chain statistics through a spatially-variable ``advective" drift in eqs. (\ref{eqn:diffusionEqn1}) and(\ref{eqn:diffusionEqn2}).

Given the partial partition functions $q$ and $q^\dagger$, the full partition function $\mathcal{Q}$ is given by
\begin{equation}
\mathcal{Q}=\int\mathrm{d}^3\mathbf{x}~q(\mathbf{x},s)q^\dagger(\mathbf{x},s). 
\label{eqn:partitionFunction}
\end{equation}
We can also calculate order parameters from $q$ and $q^\dagger$. Monomer densities (scalar order parameters) are given by
\begin{equation}
\phi_\mathrm{A}(\mathbf{x})=\frac{V}{\mathcal{Q}}\int_0^f \mathrm{d}s~q(\mathbf{x},s)q^\dagger(\mathbf{x},s)
\label{eqn:monomerDensityA}
\end{equation}
\begin{equation}
\phi_\mathrm{B}(\mathbf{x})=\frac{V}{\mathcal{Q}}\int_f^1 \mathrm{d}s~q(\mathbf{x},s)q^\dagger(\mathbf{x},s).
\label{eqn:monomerDensityB}
\end{equation}
An expression for the orientational order parameter $\mathbf{t}$ (as well as $\mathbf{Q}$) can be derived from the microscopic freely-jointed chain model (see Appendix \ref{appendix:chainStatistics}) by considering the probability that a segment of the chain at position $\mathbf{x}$ will diffuse to position $\mathbf{x}+\mathbf{r}$\cite{Zhao2012}, 
\begin{equation}
\mathbf{t}(\mathbf{x})=\frac{V}{6\mathcal{Q}}\int_0^f \mathrm{d}s\left[q\nabla q^\dagger -q^\dagger\nabla q -2\mathbf{W}qq^\dagger\right].
\label{eqn:polarOP}
\end{equation}

Given these order parameters, the mean field conditions are
\begin{equation}
w_{\mathrm{A}}(\mathbf{x})=\chi \phi_{\mathrm{B}}(\mathbf{x})+\xi(\mathbf{x})
\label{eqn:scalarMeanFieldA}
\end{equation}
\begin{equation}
w_{\mathrm{B}}(\mathbf{x})=\chi \phi_{\mathrm{A}}(\mathbf{x})+\xi(\mathbf{x})
\label{eqn:scalarMeanFieldB}
\end{equation}
where $\xi(\mathbf{x})$ is a Lagrange multiplier used to enforce the incompressibility constraint $\phi_\mathrm{A}+\phi_\mathrm{B}=1$. The vector mean field is given by 
\begin{equation}
   \mathbf{W}(\mathbf{x})=A\mathbf{t}(\mathbf{x})+C|\mathbf{t}(\mathbf{x})|^2\mathbf{t}(\mathbf{x})-K_1\nabla(\nabla\cdot\mathbf{t})-K_2\nabla\times(\nabla\times\mathbf{t})+2K_2q_0\nabla\times\mathbf{t}.
   \label{eqn:vectorMeanField}
\end{equation}
These expressions for the mean fields are obtained by finding stationary points of the free energy (eqn. \ref{eqn:fullPolarFreeEnergy}) with respect to variations in the order parameter fields, e.g. from the functional derivative $\delta F/\delta\mathbf{t}(\mathbf{x})=0$, which leads to $W_i=n\rho_0^{-1}\delta F_\mathrm{p}/\delta t_i(\mathbf{x})$.

Eqns. \ref{eqn:diffusionEqn1}-\ref{eqn:vectorMeanField} are solved by finding self-consistent values of the fields $w_\mathrm{A}$, $w_\mathrm{B}$, and $\mathbf{W}$, i.e. values which when inserted into eqns. \ref{eqn:diffusionEqn1} and \ref{eqn:diffusionEqn2}, return the same values upon calculating eqns. \ref{eqn:scalarMeanFieldA}-\ref{eqn:vectorMeanField}. This is achieved via an iterative process described in the next section.

Here we briefly summarize how the theory can also be extended to account for nematic orientational interactions, considering for specificity  the case with nematic orientational interactions, without polar interactions. Here the per chain enthalpy associated with local nematic ordering may be written as the standard fourth order expansion~\cite{deGennes} in products of the symmetric-traceless order parameter $Q_{ij} ({\bf x})$,
\begin{equation}
   F_{\mathrm{n}}^\mathrm{local}=\frac{N}{V}\int\mathrm{d}^3\mathbf{x}\left[A\mathrm{tr}(Q^2)+\frac{2B}{3}\mathrm{tr}(Q^3)+C\mathrm{tr}(Q^4)\right].
   \label{eqn:tensorLocalEnthalpy}
\end{equation}
where $Q^2 = Q_{ij}Q_{ji}$, $Q^3 = Q_{ij}Q_{jk}Q_{ki}$ and $Q^4 = Q_{ij}Q_{jk}Q_{kl}Q_{li}$.  Likewise, the free energy costs of elastic deformations can be accounted for via the second-order expansion of gradients of the nematic tensor~\cite{Mermin1989} ,
\begin{equation}
   F_{\mathrm{n}}^\mathrm{grad}=\frac{N}{V}\int\mathrm{d}^3\mathbf{x}\left[K_1(\partial_i Q_{ij})^2+K_2(\epsilon_{ikl}\partial_k Q_{lj})^2+2K_2q_0Q_{ij}\epsilon_{ikl}\partial_k Q_{lj}\right].
   \label{eqn:tensorGradientEnthalpy}
\end{equation}
In this model, the mean field effect of inter-segment orientational forces takes the form of a self-consistent tensor field, $W_{ij}({\bf x})=n\rho_0^{-1}\delta F_\mathrm{p}/\delta Q_{ij}(\mathbf{x})$, which couples to $Q_{ij}$. 
The entropy is 
\begin{equation}
S=\ln{\mathcal{Q}}+\frac{N}{V}\int\mathrm{d}^3\mathbf{x}\left(w_\mathrm{A}\phi_\mathrm{A}+w_\mathrm{B}\phi_\mathrm{B}+W_{ij}Q_{ij}\right)
\label{eqn:nematicEntropy}
\end{equation}
And the full free energy is
\begin{equation}
F=F_{\mathrm{FH}}+ F_{\mathrm{n}}^\mathrm{local}+ F_{\mathrm{n}}^\mathrm{grad}-TS.
\label{eqn:nematicEnergyFull}
\end{equation}
Based on this free energy, one can derive a SCF theory for BCPs with nematic interactions in the same spirit as the polar theory above, with the diffusion equation modified by an anisotropic diffusivity of A-segments proportional to $W_{ij}({\bf x})$, eq. (\ref{eq:dqdsQ}), and the mean order parameter $Q_{ij}$ deriving from the combined effect of the local mean-field anisotropy ($W_{ij}({\bf x}$) and symmetric, traceless second derivatives of $q$ and $q^\dagger$, eq. (\ref{eq:Q}).

As in the FFH model, the fields of the polar SCF theory can be rescaled to remove explicit $N$ dependence from the equations (see appendix \ref{appendix:rescaledSummary}). The rescaled variables are: $\bar{\mathbf{x}}=\mathbf{x}/N^{1/2}a$, $\bar{\nabla}=N^{1/2}a\nabla$, $\bar{\mathcal{Q}}=\mathcal{Q}/N^{3/2}a^3$, $\bar{V}=V/N^{3/2}a^3$, $\bar{\chi}=\chi N$, $\bar{\omega}=\omega N$, $\bar{\mathbf{t}}=N^{1/2}\mathbf{t}$, $\bar{\mathbf{W}}=N^{1/2}\mathbf{W}$, $\bar{K}=K/Na^2$, $\bar{A}=A$, and $\bar{C}=C/N$. For the nematic SCF theory, however, it is not possible to remove the independent dependence fo the theory on $N$, though we can define a similar set of rescaled variables: $\bar{Q}_{ij}=NQ_{ij}$, $\bar{W}_{ij}=NW_{ij}$, $\bar{A}=A$, $\bar{B}=B/N$, $\bar{C}=C/N^2$. Unlike the polar theory, $N$ appears explicitly in the rescaled modified diffusion equation for the nematic model, eq. (\ref{eq:dqdsQ}) and mean-field entropy eq. (\ref{eq:SQ}) .  The explicit $N$-dependence in the nematic theory derives ultimately from the fact that the nematic order parameter, as a second moment of the orientation distribution of segments, exhibits a stronger dependence on the microscopic Kuhn length as compared to the polar order parameter, which is the first moment of the segment orientation, and hence, the chain averaged effect of these higher moment orientational fluctuations exhibits a stronger $N$ dependence~\footnote{Such an effect can be seen considering the moments of segment orientation, $\langle \hat{r} \rangle$ and $\langle (\hat{r})_i (\hat{r})_j \rangle$, for chain held at fixed tension, with fixed contour length $Na$, but with variable Kuhn length $a$}.  Unlike the polar theory of flexible BCP where this $N$-dependence can be captured simply in rescaled coefficients in $F_p$, it is not possible to scale out the dependence on the microscopic length ``ultraviolet" limit of $a \to 0$ and $N \to \infty$, without completely decoupling the chain statistics from self-consistent nematic interactions. For the remainder of this article, we focus our analysis on the mean-field SCF solutions of the polar model of flexible diblocks.

\subsection{Pseudospectral implementation}

The principal computational effort of implementing the above SCF theory is in solving the diffusion equations \ref{eqn:diffusionEqn1} and \ref{eqn:diffusionEqn2}. We adapt the pseudospectral method introduced for SCFT for BCP melts by Rasmussen and Kalosakas\cite{Rasmussen2002} to incorporate the advective contributions from the self-consistent torque fields $\mathbf{W}({\bf x})$. 

An exact solution to \ref{eqn:diffusionEqn1} along the A block of the chain is given by
\begin{equation}
   q(\mathbf{x},s+\delta s)=\exp{\left[\delta s\left(\frac{N}{6}\left[a\nabla+\mathbf{W}(\mathbf{x})\right]^2-w(\mathbf{x})\right)\right]}q(\mathbf{x},s)
\label{eqn:diffusionEqnExact}
\end{equation}
where $\mathbf{W}$ and $w$ are defined to depend on the monomer species at $s$ (i.e. for the diblock model analyzed here $\mathbf{W}$ for $s >f$). Where $\mathbf{W}=0$ as in SCF treatments of the FFH model, the standard pseudospectral treatment applies: the exponential evolution operator can be approximated to second order in $\delta s$ as
\begin{equation}
   \exp{\left[\delta s\left(\frac{Na^2}{6}\nabla^2-w(\mathbf{x})\right)\right]} = \exp{\left(-\frac{\delta s}{2}w\right)} \exp{\left(\delta s\frac{Na^2}{6}\nabla^2\right)} \exp{\left(-\frac{\delta s}{2}w\right)} + \mathcal{O}(\delta s^3).
   \label{eqn:splitOperatorOld}
\end{equation}
This expansion is possible because the arguments of the exponential can linearly decomposed into terms that are either purely diagonal in real-space and those diagonal in Fourier-space (i.e. derivative) terms. This expression can be evaluated from right to left by doing the real-space multiplication $\exp{\left(-\delta sw(\mathbf{x})/2\right)}q(\mathbf{x},s)$, doing a Fourier transform and applying the operator $\exp{\left(-\delta sNa^2\mathbf{k}^2/6\right)}$ to each Fourier component with wavevector $\mathbf{k}$, and finally inverse-Fourier transforming and performing the final multiplication in real space. Due to the $\mathcal{O}(\delta s^3)$ correction, a larger finite increment for $\delta s$ can be chosen than for strictly real-space methods (e.g., Crank-Nicolson, Runge-Kutta), and this operator can be applied repeatedly to integrate the diffusion equation and solve for $q$.

When $\mathbf{W}\ne0$ and varies spatially, the expansion of the exponential operator is not so straightforward due to mixed real- and Fourier-space terms in the operator, specifically the advective operator  $\mathbf{W}(\mathbf{x})\cdot \nabla$, which is in general neither diagonal in Fourier-, nor real-space. Let us rewrite eqn. \ref{eqn:diffusionEqnExact} as
\begin{equation}
   q(\mathbf{x},s+\delta s)=\exp{\left[\delta s\left(\frac{Na^2}{6}\nabla^2+\frac{Na}{3}\mathbf{W}(\mathbf{x})\cdot\nabla+w^\prime(\mathbf{x})\right)\right]}q(\mathbf{x},s)
\label{eqn:diffusionEqnExact2}
\end{equation}
where
\begin{equation}
   w^\prime = \frac{Na}{6}\nabla\cdot\mathbf{W}+\frac{N}{6}\mathbf{W}^2-w
\label{eqn:wPrime}
\end{equation}
The operator splitting technique used in eqn. \ref{eqn:splitOperatorOld} can still be applied to separate out the purely real operator:
\begin{equation}
   q(\mathbf{x},s+\delta s)=\exp{\left(\frac{\delta s}{2}w^\prime\right)}  \exp{\left[\delta s\left(\frac{Na^2}{6}\nabla^2+\frac{Na}{3}\mathbf{W}(\mathbf{x})\cdot\nabla\right)\right]}\exp{\left(\frac{\delta s}{2}w^\prime\right)}q(\mathbf{x},s)+\mathcal{O}(\delta s^3)
\label{eqn:diffusionEqnExact3}
\end{equation}
To treat the remaining Fourier and mixed real/Fourier terms, we expand the exponential as a Tailor series to second order in $\delta s$. In doing so, it is straightforward to show that 
\begin{equation}
   \exp{\left[\delta s\left(\frac{Na^2}{6}\nabla^2+\frac{Na}{3}\mathbf{W}(\mathbf{x})\cdot\nabla\right)\right]} = \left[ 1+ U_i(\mathbf{x})\partial_i + V_{ij}(\mathbf{x})\partial_i\partial_j \right]\exp{\left(\delta s\frac{Na^2}{6}\nabla^2\right)}+\mathcal{O}(\delta s^3)
 \label{eqn:mixedOperator}
\end{equation}
where
\begin{equation}
   U_i=\frac{a\delta s}{3}W_i + \frac{a^2\delta s^2}{18}W_j\partial_jW_i + \frac{a^3\delta s^2}{36}\partial_j\partial_jW_i
 \label{eqn:aDef}
\end{equation}
and
\begin{equation}
   V_{ij}=\frac{a^2\delta s^2}{18}W_iW_j+\frac{a^3\delta s^2}{18}\partial_iW_j.
 \label{eqn:bDef}
\end{equation}
The corresponding equations for $q^\dagger$ are obtained by substituting $q \to q^\dagger$, $\partial/\partial s \to -\partial/\partial s$, and $\mathbf{W} \to -\mathbf{W}$.

The advantages of expressing the operator in this form are two-fold. Firstly, we have successfully split the operator into real- and Fourier-space operators: all derivative operators are to the right, and real space operators are to the left. Second, we have retained the $\exp{(\delta sNa^2\nabla^2/6)}$ operator, which enhances the stability of the integration scheme. It acts like a high-${\bf k}$ filter, suppressing the magnitude of high-wavenumber Fourier components.

Armed with this integration scheme for the diffusion equations, we implement a version of the software PSCF\cite{Aroraa} which we have modified to include the flexible-chain, polar SCF equations introduced in the previous section. All derivatives are calculated in Fourier space (via FFT). We use a simple mixing scheme to solve the self-consistency equations: for a given iteration $i$, input values $w_{\mathrm{in}}^i$ and $\mathbf{W}_{\mathrm{in}}^i$ for the self-consistency fields yield output values $w_{\mathrm{out}}^i$ and $\mathbf{W}_{\mathrm{out}}^i$, and the input field for the next iteration is given by
\begin{equation}
   w_{\mathrm{in}}^{i+1} = (1-\lambda)w_{\mathrm{in}}^i + \lambda w_{\mathrm{out}}^i
 \label{eqn:simpleMixing}
\end{equation}
and similarly for $\mathbf{W}$. $\lambda$ is a mixing parameter, we typically set this to 0.1.

The self consistency equations are solved for fixed unit cell dimensions; the equilibrium unit cell size (i.e. the lamellar period) $D$ can be found by optimizing the free energy of SCF solutions with respect to the unit cell size via Brent's method\cite{numerical_recipes}.

We perform our calculations using a spatial grid with $N_x=64$ grid points, and Fourier components $k_n=2\pi n/D$ where $n$ is an integer ranging from $-N/2$ to $N/2$. In solving the diffusion equations, we use an arclength step size of $\delta s=0.005$. We observe that the the SCF iterations can become unstable in high wavenumber components, depending on the highest-$k$ value needed to resolve spatial gradients. To alleviate this, we truncate high wavenumber components $|k|>\pi N/2D$ of the input $w$ and $\mathbf{W}$ fields at the beginning of each iteration.  This truncation the high-$k$ does not alter the ability to resolve self-consistency relations to required precision for the thermodynamic range explored here.

\section{Results and discussion}

\begin{figure}
\includegraphics[]{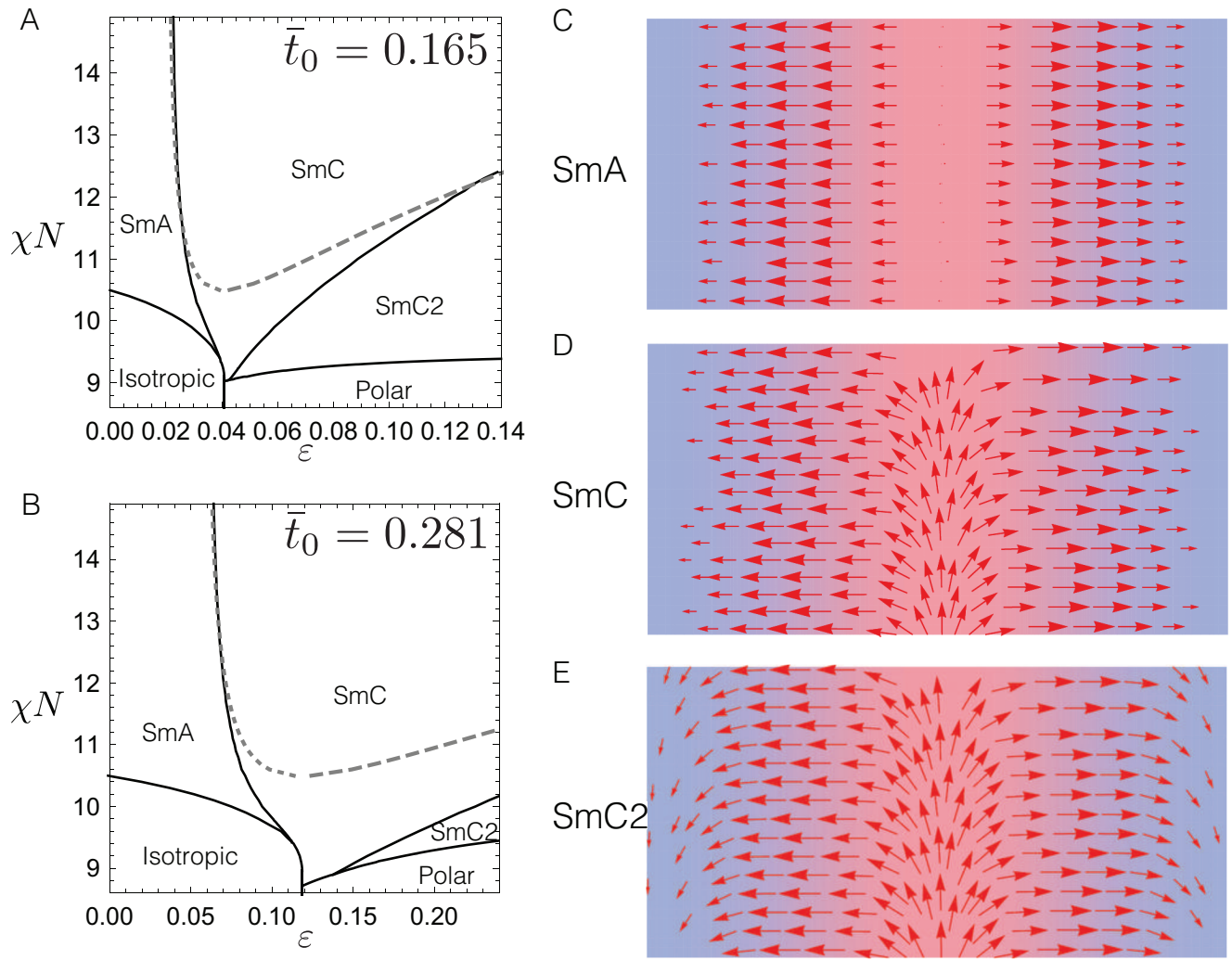}
\caption{A) Phase diagram for diblock melts with $\bar{t}_0=0.165$, for varying $\varepsilon$ and $\chi N$. As $\varepsilon$ is increased, SmC phases with tilted subdomains form. Near $\varepsilon_c$, lamellae are stabilized at lower $\chi N$. Gray lines show the predicted SmA-SmC (dotted) and SmC-SmC2 (dashed) phase boundaries based on the monomer density profiles in the FFH limit. B) Phase diagram for $\bar{t}_0=0.281$. For higher $\bar{t}_0$, the isotropic-polar transition occurs at a higher $\varepsilon_c$ value. C-E) Stream plots of $\mathbf{t}$ for SmA, SmC, and SmC2 phases overlaid on monomer density. The magnitude of $\mathbf{t}$ is indicated by streamline opacity.}
  \label{fig:phaseDiagram}
\end{figure}

We constrain our focus to structures with 1D periodicity; in other words, lamellae and disordered melts, and we fix the monomer A fraction to $f=0.5$ for which lamellae are stable morphologies the FFH model. To further simplify our analysis, we fix $\bar{t}_0$ and consider variable $\chi N$ and $\varepsilon$. In fig. \ref{fig:phaseDiagram}, we show phase diagrams for values of preferred alignment $\bar{t}_0 = 0.165$ and $0.281$, which for comparison are equal to the maximal values of polar segment orientation achieved in lamella for $\chi N=12$ and $\chi N=15$, respectively, for $\varepsilon=0$ (i.e. in the absence of self-aligning interactions).

At $\varepsilon=0$ we see the expected behavior for FFH BCPs: above a critical $\chi N=10.495$, SmA lamellar phases form, with polar segment ordering normal to the lamellar layers throughout the layer, but with the variation of sign associated with the net outward splay from A to B domains in the double layer. Below this, no phase separation occurs, and the segment ordering is isotropic. As $\varepsilon$ is raised above 0, the SmA-isotropic phase boundary occurs at lower $\chi N$ --- the tendency for local alignment stabilizes lamella. This is consistent the increased polar ordering induced by phase separation, whose additional enthalpic gain as described by $F_{local}[{\bf t}({\bf x})]$, enhances the stability of lamella relative to the isotropic disordered melt. For higher $\bar{t}_0$, lamella are stabilized at slightly lower $\chi N$: for $\bar{t}_0=0.165$, the minimum $\chi N$ that produces stable lamella is $\chi N=9.028$, and for $\bar{t}_0=0.281$, the minimum $\chi N$ that produces stable lamella is $\chi N=8.770$.

For larger $\varepsilon$, a host of interesting new phase behavior occurs. For low $\chi N$ and high $\varepsilon$, a spatially disordered homogeneous phase with uniform and spontaneously aligned polar order occurs, $\bar{t} \neq 0$.   At higher $\chi N$, we still see phase separation into lamellar structures, but these lamella have distinct symmetries, characterized by tilting of chain segments into the in-plane directions parallel to lamellar layers. Fig. \ref{fig:phaseDiagram}C-E show illustrations of these different layered structures. Smectic A phases (SmA), the familiar phase which forms in FFH BCP melts, exhibits no tilting of $\mathbf{t}$ away from the layer normal. In smectic C-like phases (SmC), $\mathbf{t}$ tilts away from the layer normal in the center of the A-rich region. At even larger enthalpy of aligment, we observe a second symmetry-breaking transition to a phase denoted as ``smectic C2" (SmC2) in which there is a second distinct B-rich region in which $\mathbf{t}$ also tilts. The in-plane tilt directions in the A-rich and B-rich regions can be either aligned or anti-aligned; there is no resolvable difference in free energy between these states, indicating that the tilts of the separate regions are decoupled, at least in the absence of Frank elastic terms. In subsection \ref{sec:homogeneous}, we discuss the role of entropic versus enthalpic effects which lead to the formation of the homogeneous isotropic and polar phases. In subsection \ref{sec:intradomain}, we expand on this to show how these effects lead to intradomain-phase transitions in the presence of compositions gradients.

We note that a qualitatively similar phase diagram is seen by Netz and Schick\cite{Netz1996}. Their study of worm-like BCPs with polar segment interactions, in a relatively flexible regime, revealed an isotropic-lamellar transition for zero orientational interaction strength (modified from the FFH theory by chain stiffness) and an isotropic-polar transition at low segregation strengths, as we find here. For stronger segregation strengths and orientational interactions, they predict what they refer to as a ``ferroelectric lamellar" phase which, similar to the SmC and SmC2 phases found here, exhibit broken symmetry and net orientational order. However, Netz and Schick note that the transition between lamellar and ferroelectric lamellar phases is discontinuous, whereas we observe a continuous transition between SmA and SmC phases. The discontinuous nature of the transition may indicate that the ferroelectric lamellar phase has the symmetry of consecutively stacked AB monolayers; i.e. the A-rich phase consists of A-blocks extending from the right side of the A-domain, and all chains are oriented with the A-block to the left and the B-block to the right. The structures observed in our study are all bilayers, i.e. the structure is made of diblocks with alternating B-A and A-B orientation, and the A-rich phase consists of A-blocks extending from the right and left sides of the A-domain. Attempts to find a self-consistent solution with the ``ferroelectric lamellar" symmetry in our theory yield convergence to the double layer symmetries described above, implying that this phase is not even metastable in our model for the regime of phase space analyzed here.   This may be attributed to the additional presence of orientation interactions in the Netz and Schick model, or possibly to the weak amount of intra-chain stiffness in their study.  More likely, however, this discrepancy arise due to the neglect of director configurations that tip out of the 1D line of symmetry considered in their study, but which are possible in self-consistent solutions described here.

\subsection{Homogeneous melts: Entropic versus enthalpic effects}\label{sec:homogeneous}

\begin{figure}
\includegraphics[]{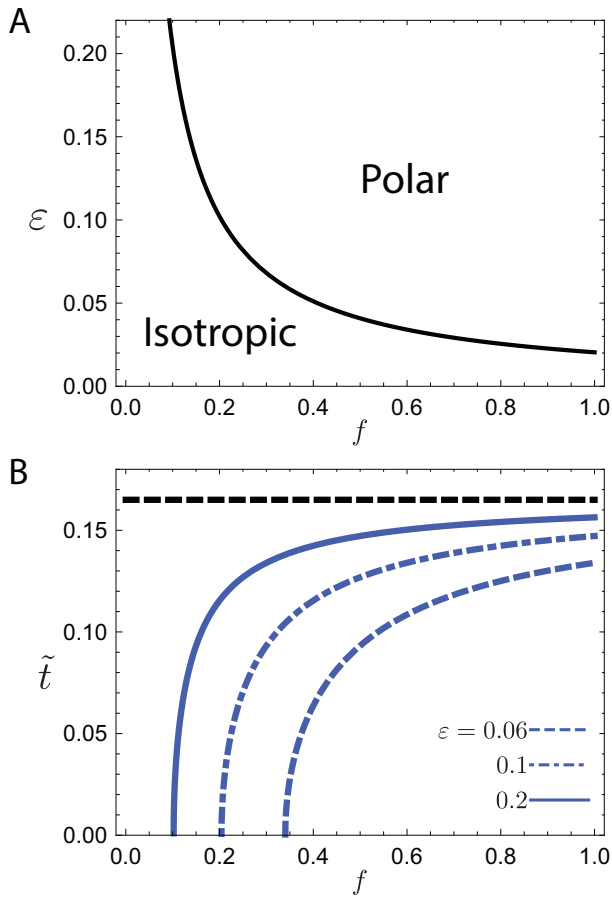}
  \caption{A) Phase diagram for homogeneous melts ($\chi N=0$) with $\bar{t}_0=0.165$ --- an isotropic phase occurs at low $f$ and low $\varepsilon$, while for higher $f$ and $\varepsilon$, melts exhibit polar order. B) The entropically modified $\tilde{t}$ for varying $f$ and $\varepsilon$. For polar states, the observed value of $|t|$ is lower than the enthalpically preferred minimum.}
  \label{fig:homogenousModifiedT}
\end{figure}

To begin understanding the role of orientational interactions in microphase segregated states shown in Fig.~\ref{fig:phaseDiagram} we consider first the spatially homogeneous state, where the monomer density $\phi_\mathrm{A}=f$ is constant, which are stable at small $\chi N$.   The orientational enthalpy (eqn. \ref{eqn:polarEnthalpy}) favors segment alignment, which competes with the entropic preference for isotropic segment distributions. Since spatial gradients vanish in the homogeneous case, it is trivial to integrate eqn. \ref{eqn:diffusionEqn1}; applying eqn. \ref{eqn:partitionFunction} and keeping only contributions from the vector fields, the entropy of eqn. \ref{eqn:polarEntropy} simplifies to
\begin{equation}
S^*=-\frac{3}{2}\frac{t^2}{f}.
\label{eqn:homogeneousVectorEntropy}
\end{equation}
For a given $t$, the entropy is inversely proportional to $f$. Intuitively, the more monomer A segments are present, the more ways a given order parameter magnitude can be achieved.

We can directly calculate the effect this has on solutions to the self-consistent field equations. Given the lack of gradients (i.e. $\nabla q=\nabla q^\dagger=0$), eqn. \ref{eqn:polarOP} simplifies to 
\begin{equation}
\mathbf{t}(\mathbf{x})=-\frac{f}{3}\mathbf{W}(\mathbf{x})
\label{eqn:tH}
\end{equation}
and combined with eqn. \ref{eqn:vectorMeanField} results in the self-consistency condition
\begin{equation}
0=\left(\frac{3}{f}-\frac{4\varepsilon}{t_0^2}\right)\mathbf{t}+\frac{4\varepsilon}{t_0^4}|\mathbf{t}|^2\mathbf{t}
\label{eqn:homogeneousSelfConsistency}
\end{equation}
which has solutions
\begin{equation}
\tilde{t}(f)=\begin{cases}
    t_0\sqrt{1-\frac{3}{4}\frac{t_0^2}{\varepsilon f}} & \text{if $\varepsilon>\frac{3}{4}\frac{t_0^2}{f}$}.\\
    0 & \text{otherwise}.
  \end{cases}
\label{eqn:tEntropyMod}
\end{equation}
Thus, due to the rotational entropy of chain segments, homogeneous melts only exhibit polar order above a critical $\varepsilon_c=\frac{3}{4}\frac{t_0^2}{f}$, as opposed to a system which is governed only by the enthalpic interactions described by eqn. \ref{eqn:polarEnthalpy} which exhibits polar order for all $\varepsilon>0$. When melts do exhibit polar order, this alignment is slightly reduced below the enthalpically preferred value $t_0$. Fig. \ref{fig:homogenousModifiedT}A illustrates the phase diagram for homogeneous melts with fixed $t_0$. The polar phase is stable only at high $\varepsilon$ and $f$; at low $\varepsilon$ and $f$, entropy dominates. Fig. \ref{fig:homogenousModifiedT}B illustrates the $f$-dependence of the entropic cost to alignment --- the equilibrium polar order $\tilde{t}$ is always lower than $t_0$, but it approaches $t_0$ when $f=1$ at high $\varepsilon$. The boundaries between the isotropic and polar phase in fig. \ref{fig:phaseDiagram}A and B agree with this prediction, $\varepsilon_c=\frac{3}{4}\frac{t_0^2}{f}$ at low $\chi N$. 

\subsection{Intradomain phase transitions}\label{sec:intradomain}

\begin{figure}
\includegraphics[]{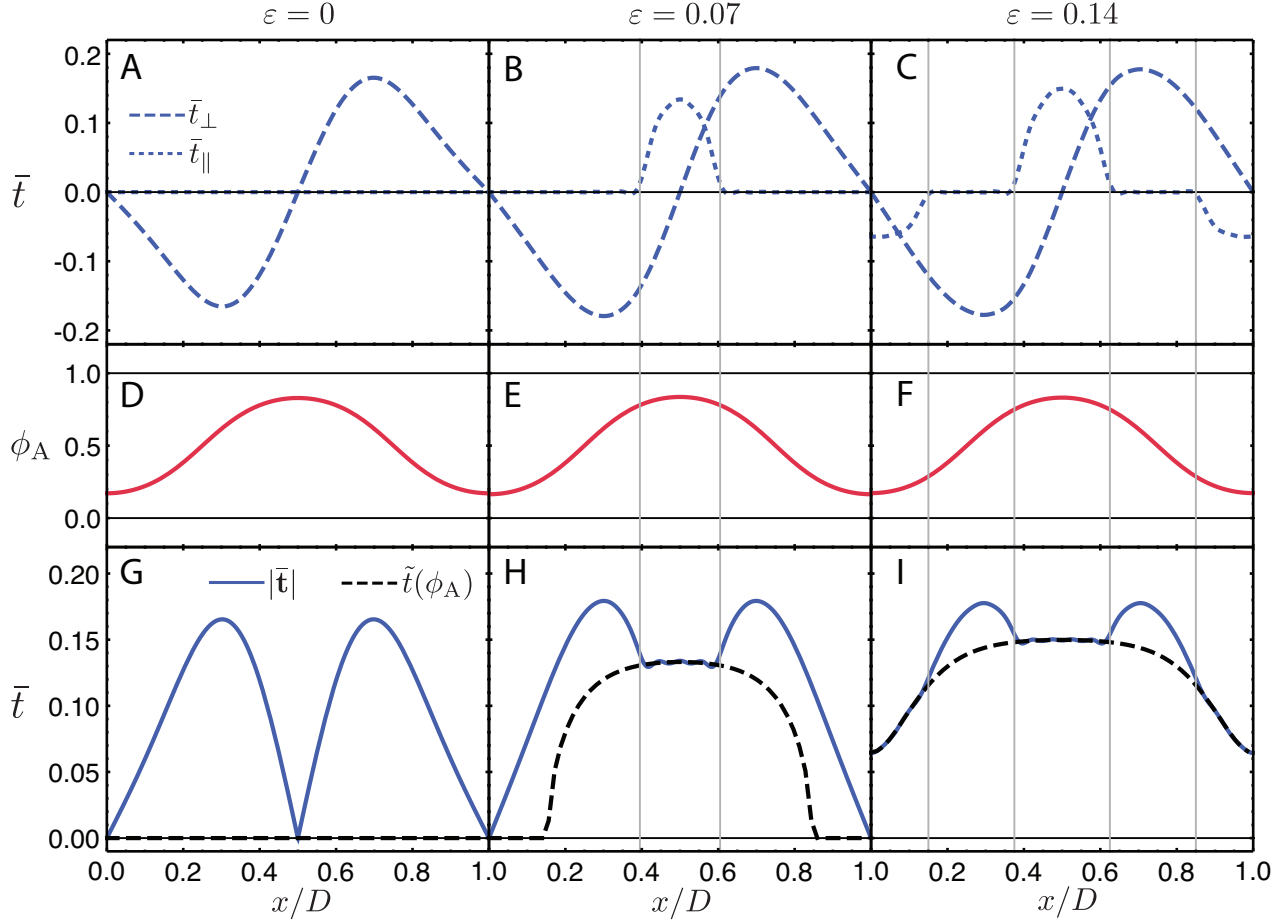}
\caption{A-C) Components of $\bar{\mathbf{t}}$ perpendicular and parallel to the lamellar layers for  $\epsilon=0.0$ (A), $\epsilon=0.07$ (B), and $\epsilon=0.14$ (C), in a melt with $\chi N=12$, $f=0.5$, and $\bar{t}_0=\mathrm{max}(\bar{t}_{\mathrm{FFH}})=0.165$. As $\epsilon$ increases, discrete sections of the melt exhibit a parallel component. Vertical lines indicate boundaries between tilted and non-tilted phases. (D-F) The corresponding plots of the density profile $\phi_\mathrm{A}$. The density profile is largely insensitive to $\varepsilon$. (G-I) The corresponding plots of $|\bar{\mathbf{t}}(x)|$ and $\tilde{t}(\phi_\mathrm{A}(x))$. Non-zero $\bar{t}_\parallel$ occurs where $|\bar{\mathbf{t}}|=\tilde{t}(\phi_\mathrm{A})$ due to spontaneous symmetry breaking. As $\varepsilon$ increases, symmetry breaking occurs first where $\phi_\mathrm{A}$ is highest.}
  \label{fig:tProfiles}
\end{figure}

We now turn our attention to lamellar phases and consider the coupling between spatial gradients and segment director.  One might anticipate two main effects that density variations will have. First, even in the absence of orientational interactions (i.e. the FFH model), monomer density variations in phase separated BCPs lead to orientational order which couples to the domain geometry~\cite{Prasad2017}. Second, because of the density-dependence of the thermodynamics of alignment, as discussed in subsection \ref{sec:homogeneous}, the relative susceptibility of A segments to the enthalpic drive to align will vary throughout the unit cell with composition.

Fig. \ref{fig:tProfiles}A-C show the components of the $\mathbf{t}$ field parallel and normal to the lamellar layers for SmA, SmC, and SmC2 phases, respectively, and correspond to the profiles in fig. \ref{fig:phaseDiagram}C-E. We see that when $\mathbf{t}$ exhibits tilt into the parallel direction in SmC and SmC2 phases, it does so only in discrete portions of the unit cell, remaining strictly normal in the rest of the cell.  In this sense, this state remains SmA-like in regions away from A-rich core, with the symmetry-breaking of in-plane tilt confined the central core of the domain.  

It is instructive to consider the equilibrium value of polar magnitude, $\tilde{t}({\bf x})$, that would arise in homogeneous melts with the same local A-segment composition $\phi_\mathrm{A}(\mathbf{x})$ and without consideration of the additional constraints of chain connectivity, as described by contributions from composition gradients (as in the first two terms of the integrand of eqn. \ref{eqn:polarOP}). That is, simply taking the expression eqn. \ref{eqn:tEntropyMod} and replacing composition $f$ with local A segment density, $\phi_\mathrm{A}(\mathbf{x})$ to arrive at
\begin{equation}
\tilde{t}({\bf x}) =\begin{cases}
    t_0\sqrt{1-\frac{3}{4}\frac{t_0^2}{\varepsilon \phi_\mathrm{A}(\mathbf{x})}} & \text{if $\varepsilon>\frac{3}{4}\frac{\bar{t}_0^2}{\phi_\mathrm{A}(\mathbf{x})}$}.\\
    0 & \text{otherwise}.
  \end{cases}
  \label{eq:tildex}
\end{equation}
Fig. \ref{fig:tProfiles}D-F show $\tilde{t}({\bf x})$ in comparison to the magnitude $|\bar{\mathbf{t}}|$ predicted for self-consistent solutions.  Notably, we find that $|\bar{\mathbf{t}}| =\tilde{t}({\bf x})$ in the precise regions where tilt is non-zero ($t_\parallel \neq 0$) and, elsewhere, where the director resembles profiles generated by phase separation along the lamellar normal, its magnitude $|\bar{\mathbf{t}}|$ never falls below $\tilde{t}({\bf x})$. 

Noting that the gradient terms in the self consistency relation for ${\bf t}({\bf x})$, eq. \ref{eqn:polarOP}, vanish by symmetry in the centers of the A- and B-rich domains, we can use the predictions for the isotropic to polar transition of the uniform melt to predict the onset of in-plane tilt in ordered lamella, assuming additionally that orientational interactions do not too strongly alter the monomer density profiles. Using the $\varepsilon\to 0$ limit $\phi_\mathrm{A}^\mathrm{FFH}$, we calculate for a given $\chi N$ the critical value of $\varepsilon$ at which the isotropic-polar transition occurs given the monomer density $\phi_\mathrm{A}^\mathrm{FFH}$ in the center of the A-rich phase to predict the SmA-SmC phase boundary, as well as the critical epsilon corresponding to $\phi_\mathrm{A}^\mathrm{FFH}$ at the center of the B-rich phase to predict the SmC-SmC2 boundary. The results are plotted as gray lines in fig. \ref{fig:phaseDiagram}A.  We see that at high-$\chi N$ and low-$\epsilon$, this simple prediction agrees well with self-consistent solutions. The prediction deviates from the observed phase boundaries in the opposite region of relatively small-$\chi N$ and larger-$\epsilon$, due to the increased segregation induced by strong orientational interactions relative to scalar repulsion alone.

Taken together, the behavior described above can be considered as a type of \emph{intra-domain phase transition}, in which regions of distinct symmetry and local order parameter behavior coexist in the same mesodomain of A-segments, delineated by a sharp transition between them. In one ``phase", the {\it normal phase}, the orientational order is dominated by the interplay between spatial gradients and chain connective, nominally equivalent to the alignment seen in lamellae for $\varepsilon =0$.  The other ``phase", the {\it tilt phase}, derives from the thermodynamic instability of isotropic segments in homogeneous melts, spontanenous breaking the rotational symmetry of the lamellar state around the layer normal.

To understand emergence distinct intra-domain ``phases", we reframe eqn. \ref{eqn:polarOP} to emphasize separate contributions from gradient-induced effects (deriving from chain connectivity) and enthalpy of local segment alignment:
\begin{equation}
   \mathbf{t}=\mathbf{t}^\mathrm{grad}+\mathbf{t}^\mathrm{enth}
   \label{eqn:tParts}
\end{equation}
where
\begin{equation}
   \mathbf{t}^\mathrm{grad}=\frac{V}{6\mathcal{Q}}\int_0^f \mathrm{d}s\left[q\nabla q^\dagger -q^\dagger\nabla q\right]
   \label{eqn:tGrad}
\end{equation}
and
\begin{equation}
   \mathbf{t}^\mathrm{enth}=\frac{\phi_\mathrm{A}}{3} {\bf W}= \frac{\phi_\mathrm{A}}{3}\left(A+C|\mathbf{t}|^2\right)\mathbf{t}.
   \label{eqn:tEnth}
\end{equation}
The definition in eqn. \ref{eqn:tEnth} is arrived at from combining the last term in eqn. \ref{eqn:polarOP} with eqns. \ref{eqn:vectorMeanField} and \ref{eqn:monomerDensityA}. Because $\mathbf{t}^\mathrm{grad}$ is purely normal to the lamellar layers due to the 1D periodic symmetry of the composition profile, we write eqn. \ref{eqn:tParts} as two equations for the normal and parallel components of the polar order parameter:
\begin{align} 
   t_\perp &= t^\mathrm{grad}_\perp-\frac{\phi_\mathrm{A}}{3}\left[A+C(t_\perp^2+t_\parallel^2)\right]t_\perp \\
   t_\parallel &= -\frac{\phi_\mathrm{A}}{3}\left[A+C(t_\perp^2+t_\parallel^2)\right]t_\parallel
   \label{eqn:tPartsXY}
\end{align}
These self-consistency conditions for ${\bf t}$ admit two types of solutions. The first requires $t_\parallel=0$ and the value of $t_\perp$ is slaved to $ t^\mathrm{grad}_\perp$ through a cubic polynomial.  The second occurs when $|\mathbf{t}|^2=-(A+\phi_\mathrm{A}/3)/C=\tilde{t}^2$, and requires $\mathbf{t}^\mathrm{grad}=0$. Solutions of the first kind correspond to alignment which is purely normal to the layers and closely-follows the intra-domain director patterns induced by phase separation and chain connectivity. The second solution corresponds to the regions of spontaneous symmetry breaking by in plane tilt, and moreover, the magnitude of alignment follows $|\mathbf{t}|=\tilde{t}$ predicted by the density-only condition of eq. (\ref{eq:tildex}).

\begin{figure}
\includegraphics[]{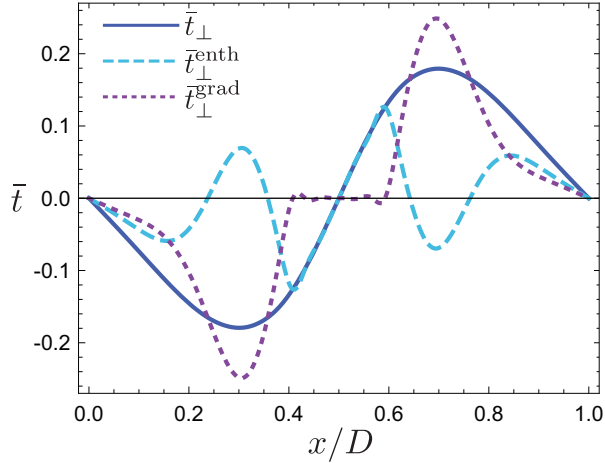}
\caption{Separate $\bar{t}^\mathrm{grad}$ and $\bar{t}^\mathrm{enth}$ contributions to $\bar{t}_\perp$, for a smectic C melt with $\chi N=12$, $\bar{t}_0=0.165$, and $\varepsilon=0.07$. As in the simplified model of eqns \ref{eqn:tPartsXY}, tilting only occurs where $\bar{t}^\mathrm{grad}=0$ --- the ordering is purely enthalpic in these regions. }
  \label{fig:tSplit}
\end{figure}

We plot the explicit contributions from $\mathbf{t}^\mathrm{grad}$ and $\mathbf{t}^\mathrm{enth}$ from our SCF solutions in Fig. \ref{fig:tSplit} shows for a SmA phase, and that $\mathbf{t}^\mathrm{grad}$ vanishes in the tilted regions where $|\mathbf{t}|=\tilde{t}$. This non-analytic behavior (e.g. discontinous first derivatives) of $\mathbf{t}^\mathrm{grad}$ profiles is consistent with the attribution of the two classes of solutions to eqns. (\ref{eqn:tPartsXY}) as distinct ``phases" of local alignment: $\mathbf{t}$ is either purely normal to the lamella, or $|\mathbf{t}|=\tilde{t}$ while $\mathbf{t}^\mathrm{grad}=0$.

One question remains: what determines the location of the intra-domain phase boundary, between $\tilde{t}=\mathbf{t}_\perp$ and $t_\parallel \neq 0$? We observe that, qualitatively, the $t_\perp$ in SCF profiles changes little as $\varepsilon$ is raised from 0, as evident in fig. \ref{fig:tProfiles} A-C. The reason derives from strong topological dependence of $t_\perp$ on the density of A-block ends. The orientational order parameter field $\mathbf{t}$ can be related to the areal flux ${\bf J}_A$ of A-segments pointing along the A-block through ${\bf J}_A({\bf x}) = \rho_0 a \mathbf{t}({\bf x})$\cite{Svensek2013}.  The divergence of this flux is related to a conservation law\cite{deGennes1976} of the local density A-block ends, $\rho_+$ minus the local density of the A-B junctions, $\rho_-$:
\begin{equation}
   \nabla\cdot {\bf J}_A=\rho_+ - \rho_-
   \label{eqn:flux}
\end{equation}
For 1D periodic solutions, this condition slaves the normal derivatives of $t_\perp$ to the spatial pattern of A-block ends, $\delta \rho ({\bf x} ) = \rho_+({\bf x} )  - \rho_-({\bf x} ) $, through $a \rho_0 \partial_\perp t_\perp = \rho_+({\bf x} )  - \rho_-({\bf x} )$.  Notably, under the weak to moderate segregation conditions considered here, we expect this relative density to vary in proportion to the local A excess, or $\delta \rho ({\bf x} ) \propto \phi_{\rm A} ({\bf x}) - f$.  Hence, to a large extent, the normal component of the polar director is slaved to the composition pattern, forced to change sign and pass through 0 twice within a given domain.  

\begin{figure}
\includegraphics[]{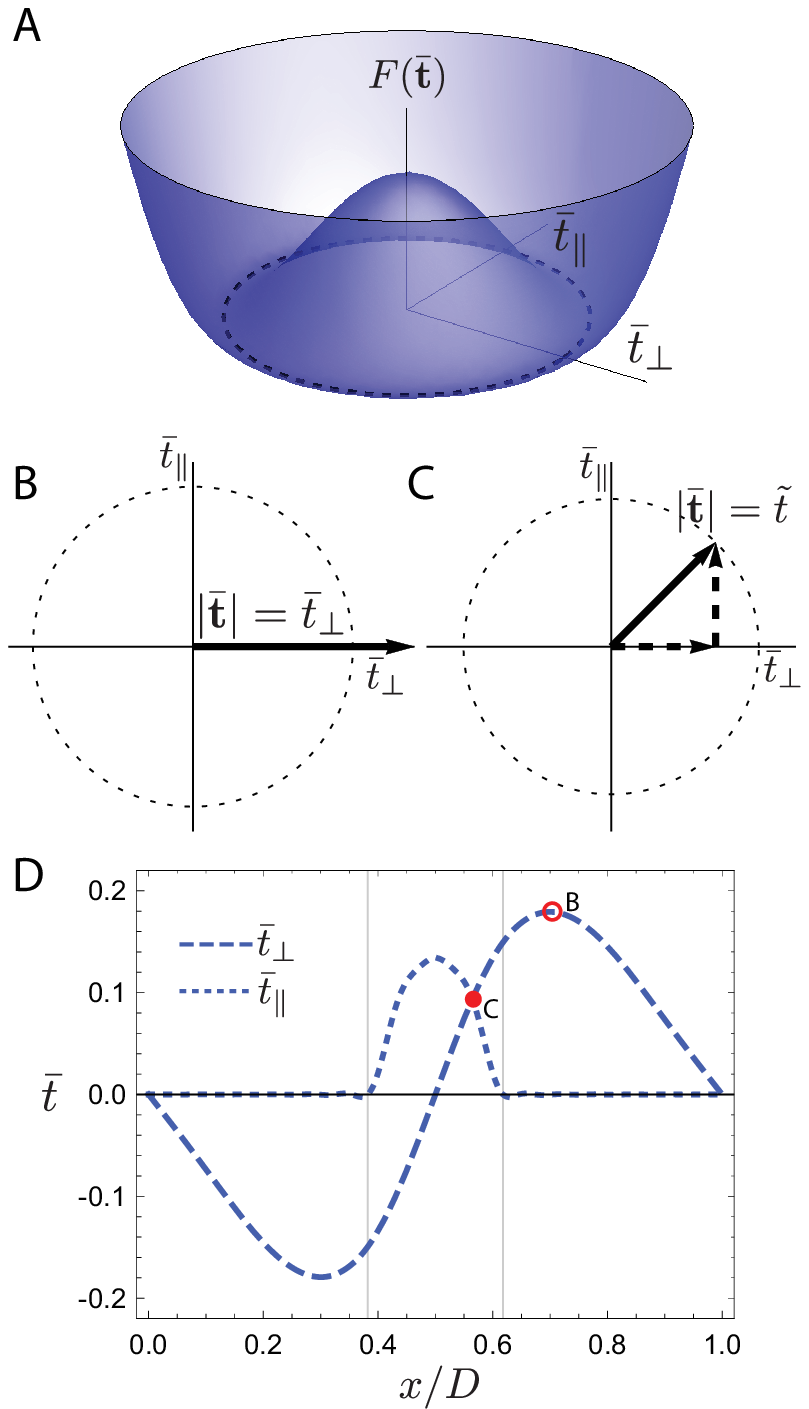}
\caption{A) The free energy landscape associated with $\mathbf{t}$ including both enthalpic and entropic contributions, shown explicitly as a function of $t_\perp$ and $t_\parallel$. The degenerate free energy minimum, indicated by a dashed circle, corresponds to $|\mathbf{t}|=\tilde{t}$.  B-C) The $t_\perp$ and $t_\parallel$ components of $\mathbf{t}$ corresponding to the points highlighted in (D). The dashed circle again corresponds to the energy minimum at $|\mathbf{t}|=\tilde{t}$. B) Where $t_\perp > \tilde{t}$, there is no way to decrease the energy by introducing a $t_\parallel$ component. C) Where $t_\perp < \tilde{t}$, the energy can be minimized by introducing a $t_\parallel$ component such that $|\mathbf{t}|=\tilde{t}$. D) The $\mathbf{t}$ profile of a SmC lamella. The open circle highlights a point with normal ordering; the closed circle highlights a point with tilted ordering.}
  \label{fig:energyLandscape}
\end{figure}

Taking $t_\perp({\bf x})$ to be largely fixed by the degree of phase segregation, the origin of the sharp ``intra-domain" phase boundary becomes clear when considering the local free energy landscape at ${\bf x}$ as a function of variable $\mathbf{t}$, shown schematically in Fig. \ref{fig:energyLandscape}A.  This effective free energy has a global equilibrium value for $|\mathbf{t}|=\tilde{t}({\bf x})$, but it also subject to the constraint of fixed values of $t_\perp({\bf x})$ imposed by polarization of chain ends.  Consider the points in regions where $|\mathbf{t}|=t_\perp$, and where $|\mathbf{t}|=\tilde{t}$, as highlighted in fig. \ref{fig:energyLandscape}B.  When in regions where phase separation leads to $|t_\perp|>\tilde{t}$, the effective free energy is minimized for $t_\parallel = 0$ (fig. \ref{fig:energyLandscape}D), while in regions where $|t_\perp|<\tilde{t}$ (fig. \ref{fig:energyLandscape}D), the effective free energy is unstable at $t_\parallel = 0$ as the polar director tilts up to $t_\parallel \neq 0$ values that maintain the minimal free energy along the circle of constant $|{\bf t}| = \tilde{t}$.  In this phase transition description, the value of $t_\perp({\bf x})$ induced by polarization of A-block ends and the local composition $\phi_{\rm A} ({\bf x})$ acts as a thermodynamic control variable within the given mesodomains, describing the continuous variation of the local susceptibility to of weakly oriented segments to spontaneously alignment.

\subsection{Domain spacing}

\begin{figure}
  \includegraphics[]{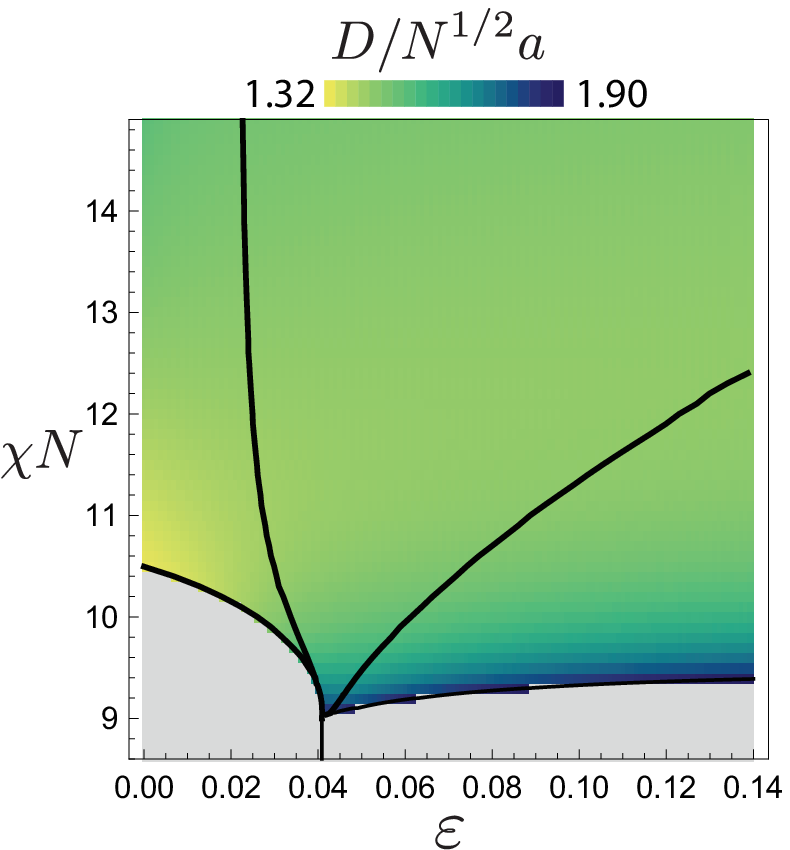}
  \caption{Lamellar period as a function of $\chi N$ and $\varepsilon$ for $\chi N=12$ and $t_0=0.165$. Near the SmC2-Polar boundary, the unit cell grows to \textasciitilde$2N^{1/2}a$}
  \label{fig:periodDiagram}
\end{figure}

Fig.~\ref{fig:periodDiagram} shows a map of the equilibrium lamellar period $D$ in th  $\chi N$-$\varepsilon$ for the case of preferred alignment, $t_0=0.165$.  For relatively small enthalpies of alignment, as in the FFH theory, $D$ is shown to increase monotonically with segregation strength $\chi N$.  For larger $\varepsilon$, above the isotropic-to-polar transition point of the homogeneous melts, we observed an anomolous dependence of domain spacing on segregation, in which $D$ increases as $\chi N$ {\it decreases} towards the SmC2-Polar phase boundary.  Along this boundary layer spacing reaches $D \simeq 1.9 N^{1/2}a$, a nearly 50\% increase over the is value ($D \simeq 1.32  N^{1/2}a$) at the ODT for $\varepsilon = 0$.

This anomalous behavior can be explained by considering that interplay between the inter-layer compression elasticity of the lamella and the enthalpy gain of enhanced alignment in stretched domain.  Following a standard approach to describe the generalized elasticity of smectic layers near second-order onset of periodic spatial order\cite{chaikin2000principles}, we estimate the (per chain) elastic cost to stretch layers from a preferred spacing $D_0$ to a larger value $D$ as $F_{layer} \approx B \lambda^2/2$, where $\lambda=(D-D_0)/D_0$ is the layer strain and the bulk modulus $B \propto |\delta \phi|^2$ grows with mean value of the density modulation order parameter.  Here we take $D_0$ to be the spacing in the absence of enthalpy of alignment, which can be written in the form $F_{local} = \varepsilon (|{\bf t}|^2 - t_0^2)^2$.  For the normal oriented segments, the (mean) polar order couples to domain spacing according to $\langle t_\perp \rangle = D/(N a) \approx D_0 (1+ \lambda)/(N a)$.  Combining these two effects and considering relative small layer dilations, we find
\begin{equation}
F(\lambda) \approx \frac{B}{2} \lambda^2 -\varepsilon \frac{t_0^2}{N ^{1/2}} \lambda ,
\end{equation}
where we used $D_0 \approx  N^{1/2}a$.  Minimizing over layer strain, we predict equilibrium  layer spacing
\begin{equation}
D \approx D_0 + \frac{ \varepsilon t_0^2}{N^{1/2} B} .
\end{equation}
Hence, as segregation strength decreases towards the homogeneous state $\delta \phi \to 0$ and the bulk modulus drops, the layer spacing between increasing compliant to elastic stretching of the layers favored enthalpic alignment of segments.  

\subsection{Frank elasticity effects}

\begin{figure}
\includegraphics[]{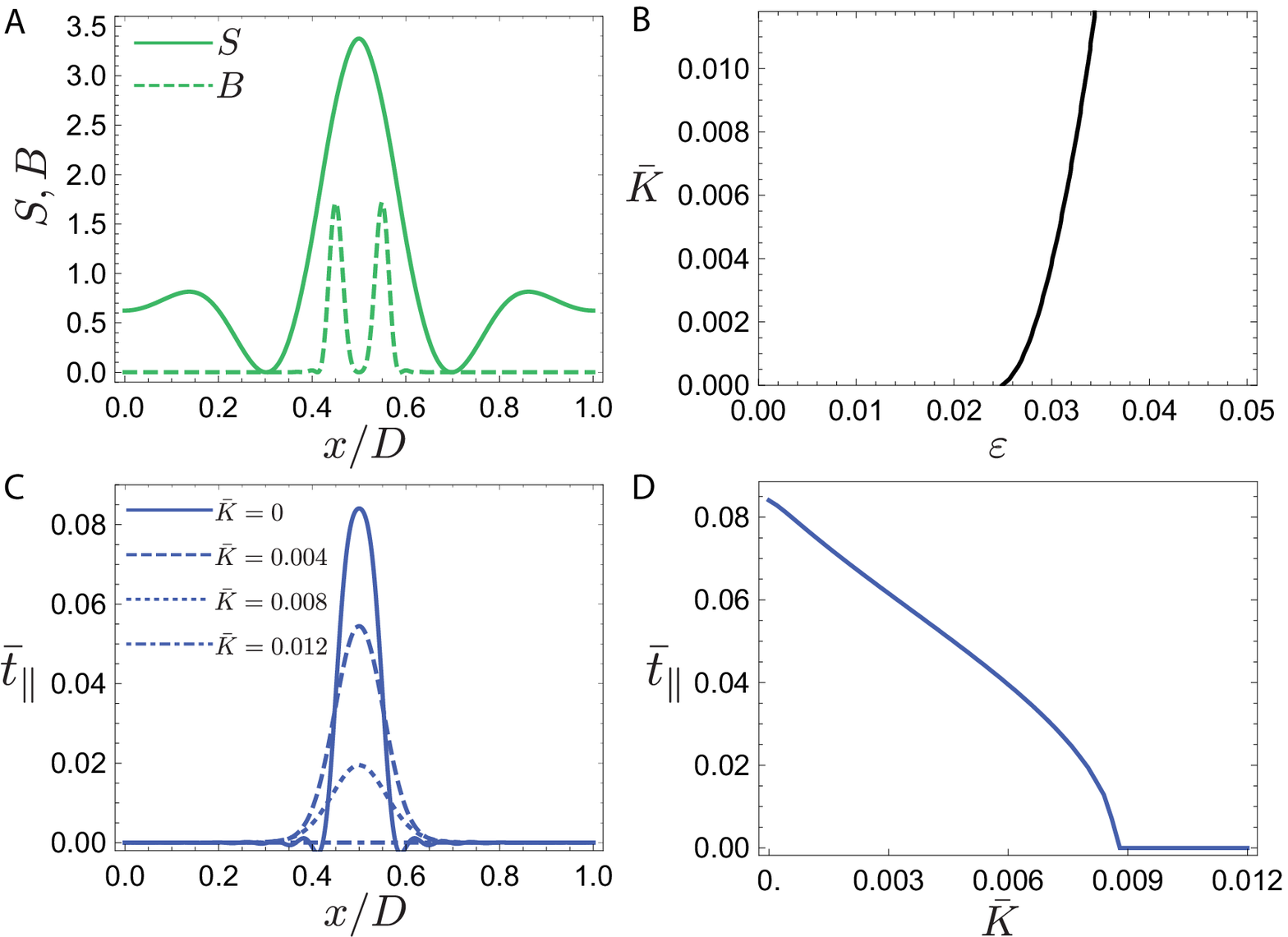}
  \caption{A) Splay (solid) and bend (dashed) profiles for  SmC lamella with $\varepsilon=0.033$ and $\chi N=12$. In the tilted subdomain, SmC phases exhibit bend. B) SmA-SmC phase boundary in the $\bar{K}$-$\varepsilon$ plane, for $\chi N=12$ and $\bar{t}_0=0.165$. As elastic costs are increased, higher $\varepsilon$ is needed to induce a SmA-to-SmCtransition. C) $t_\parallel$ profiles for lamella with $\varepsilon=0.033$ and varying $\bar{K}$. As $\bar{K}$ increases, the magnitude of tilt decreases, and the width of the tilted region increases slightly. E) The maximum value of $\bar{t}_\parallel$ as a function of $\bar{k}$ for $\varepsilon=0.033$.}
  \label{fig:gradients}
\end{figure}

Here, we briefly consider the thermodynamic cost of director gradients in intra-domain textures, and their influence of distinct intra-domain symmetries.  The planar symmetry of the lamellar states found in the phase diagram in Fig. ~\ref{fig:phaseDiagram} generate both splay and bend deformations of ${\bf t}$ (no twist). Defining the splay- and bend-density, respectively as $S=(\bar{\nabla}\cdot\bar{\mathbf{t}})^2$ and $ B=(\bar{\nabla}\times\bar{\mathbf{t}})^2$, we analyze the director gradients in the SmC phase in fig. \ref{fig:gradients}A.  In fact, the tilted regions can be characterized as Neel ``splay walls"\cite{Ding1993}. In nematics, such splay walls tend to form when there is a greater penalty for bend than splay, i.e. $K_1<K_2$, 

To test the effect of elastic costs, we consider a one constant approximation $K_1 = K_2 = K$. Fig. \ref{fig:gradients}B shows the SmA-SmC phase boundary for fixed  $\chi N=12$ (weak segregation) and $t_0=0.165$ for varying scaled elastic constant $\bar{K}=K/(Na^2)$ and $\varepsilon$. As $\bar{K}$ is increased, the SmA-SmC boundary shifts towards higher $\varepsilon$ --- as the cost of elastic deformations of the segment director, a higher enthalpy of alignment is needed to overcome this additional elastic penalty to reorient the director in the tilted core. This is illustrated in fig. \ref{fig:gradients}C, which shows $\mathbf{t}$ profiles for $\varepsilon=0.033$ at different values of $\bar{K}$. For $\bar{K}=0$, the melt is in the SmC phase. As $\bar{K}$ is increased to 0.012, the elastic costs suppress the bend deformations associated with the SmC phase, resulting in the formation of a SmA phase. We also observe that, where the elastic costs do not completely suppress the tilt, the tilted region widens slightly due to the great elastic cost of having a non-differentiable $\bar{t}_\parallel$ profile. In fig. \ref{fig:gradients}D we show the maximum value of $\bar{t}_\parallel$ as a function of $\bar{K}$. The tilt reduces in a continuous fashion to zero.

The sequence of intra-domain alignment with increasing Frank constant is analogous to the Freedericksz transition in nematic liquid crystals\cite{deGennes}. The Freedericksz transition occurs when a nematic liquid crystal is confined between two parallel boundaries and an applied magnetic field causes a reorientation of the director away from the boundary aligned values (say homeotropic). Consider boundaries with homeotropic anchoring conditions (causing the equilibrium director field to be normal to the boundaries), and a magnetic field is applied along the plane of the boundaries. Above a critical applied field $H_c=(\pi/d)(K/\chi_m)^{1/2}$, where $d$ is the spacing between plates, $K$ is the Frank elastic constant in a one-constant approximation, and $\chi_m$ is the magnetic susceptibility, the magnetic field induces in-plane tilt in the director field, but below $H_c$ the field is purely normal to the boundaries.  Defining $\Delta \varepsilon$ as the excess enthalpy of alignment (above the $K=0$ value) to stabilize in-plane tilt an the lamellar core, we find a similar depend on SmA-SmC boudnary, $\Delta \varepsilon \sim K^{1/2}$, in the small-$K$ limit.

\section{Conclusion}

Due to the density-dependent entropic cost for polar order in BCP melts,  BCPs with a preference for local orientational order exhibit intradomain phase transitions between regions dominated by enthalpic orientational interactions and order arising from composition gradients and chain connectivity. The nature of this phase transition is most clearly demonstrated in homogeneous BCP melts: for a given degree of preferred order $t_0$ and enthalpy of ordering $\varepsilon$, there is a critical monomer fraction $f$ below which the melt is isotropic, and above which it exhibits uniform polar order. In lamellae, where there are spatial variations in monomer density, the same type of transition occurs, though the situation is more complicated due to order arising from composition gradients and chain connectivity. Where density gradients vanish due to symmetry, regions of orientational order driven purely by enthalpic effects appear if $\varepsilon$ is large enough; in these regions, the polar order parameter tilts away from the lamellar normal vector, breaking the lamellar symmetry. Where density gradients are large, however, order arising from composition gradients and chain connectivity dominates.

This discovery of intra-domain phase transitions reveals a new class of mesophase symmetries in BCPs: rather than transitions between phases with monomer density profiles of different symmetries and network connectivities, these phase transitions occur between phases with qualitatively different orientational order. In the lamellar structures studied here, we see a periodic variation in the average orientation of chain segments. This variation could be leveraged if the orientational order were associated with an optical easy axis, leading to optical metamaterials.

This work also represents an import step forward in the theoretical understanding of orientational interactions in flexible BPCs. Previous work has studied some aspects of orientational gradient penalties, including a preference for chirality; here we introduced local orientational preference and took first steps to combine these two type of orientational interactions. This paper considered diblocks in which only one of the blocks' segments interact orientationally, but the theory is generalizable to more complicated chain architectures and could include orientational interactions between segments of different monomers. Beyond this there remains a large unexplored parameter space. Local ordering is unexplored in more complicated structures such as spherical, cylindrical, and network phases. In cylindrical structures without a preference for local orientational order, the polar order goes to zero at the center of a cylinder. Including a preference for local order would likely lead to tilting along the axis of the cylinders. In gyroid networks consisting of cylinder-like struts which meet at three-fold nodes, there is likely to be a frustration in the tilt direction: the tilt in one strut will be into the node, the tilt in another will be out of the node, leaving the tilt in the third strut undetermined. The topology of the phase boundaries plus the preference for local ordering may also lead to topological defects in the orientational order; adding gradient penalties then may affect the stability of these structures. A preference for chirality combined with a preference for local order may also lead to interesting effects\cite{Grason2015a}. Previous work on chirality has considered only gradient penalties which affect the order arising from composition gradients and chain connectivity~\cite{Zhao2013,Wang2013}. Chirality in regions dominated by enthalpically preferred orientational order, such as the core tilted cores of so-called H* domains\cite{Ho2012}, may promote an even stronger coupling between inter-segment twist and mesochiral domain shapes in chiral BCP.

We introduced in this paper a theory for nematic orientational interactions, but we leave a computational study for future work. The parameter space for nematically-interacting BCPs is much larger than that of the polar case. The intra-domain textures of nematic alignment are far more complex than the polar case, even in the absence of inter-segment alignment.  Well segregated domains of the FFH model exhibit two distinct zones of alignment corresponding to calamitic (normal alignment in the brushes) and discotic ordering (in plane alignment at the AB interface), which  have different degrees of preferred order and presumably very different susceptibilities to further alignment in the presence of local self-align forces. This may lead to sub-domains with ordering corresponding to different energy minima. Between these subdomains, the nematic order must transition from one preferred direction for another, which in turn is expected to induce addtional penalties for the more complex gradients in the nematic orientation needed to reorient from on sub-domain to another. The host of new BCP phases of intra-domain nematic alignment, as well their additional strong dependence on chain length, remain to be explored.

\section*{ACKNOWLEDGEMENTS}
The authors are grateful I. Prasad for important discussions on segment order parameters and orientational SCFT and to E. Thomas for helpful comments on the work.   This work was supported by the U.S. Department of Energy, Office of Science, Office of Basic Energy Sciences, under Award No. DE-SC0014549.

\appendix

\section{Chain distributions and order parameters from microscopic chain model}\label{appendix:chainStatistics}

Using a freely jointed chain model, we can derive the diffusion equations \ref{eqn:diffusionEqn1} and \ref{eqn:diffusionEqn2} based on chain conformation statistics under the influence of the polar self consistent torque field $\mathbf{W}({\bf x})$. We begin with the transfer probability $p(\mathbf{x}',\mathbf{x})$ which gives the probability that a segment of length $a$ with one end at $\mathbf{x}$ will have its other end at $\mathbf{x}'$:
\begin{equation}
p_p(\mathbf{x}',\mathbf{x}) = \frac{\delta(|\mathbf{x}-\mathbf{x}'|-a)}{4\pi a^2}\exp{\left(-\omega(\bar{\mathbf{x}})-\mathbf{W}(\bar{\mathbf{x}})\cdot\frac{\mathbf{x}-\mathbf{x}'}{a}\right)}
\label{eqn:transferProbabilityPolar}
\end{equation}
where $\bar{\mathbf{x}}=(\mathbf{x}+\mathbf{x}')/2$ is the midpoint of the segment, and $\omega({\bf x})$ is the per segment self-consistent (scalar) chemical potential.

To derive the diffusion equation for $q(\mathbf{x},s)$, we write down the probability that segment $n$ at $\mathbf{x}-\mathbf{r}$ will diffuse to $x$:
\begin{equation}
q(\mathbf{x},n+1) = \int\mathrm{d}^3\mathbf{r}~p_p(\mathbf{x},\mathbf{x}-\mathbf{r})q(\mathbf{x}-\mathbf{r},n)
\label{eqn:microscopicDiffusionEqn}
\end{equation}
Expanding $p$ and $q$ in powers of $\mathbf{r}$, this can be written as 
\begin{eqnarray}
\nonumber
q(\mathbf{x},n+1) &= &\int\mathrm{d}^3\mathbf{r}~\frac{\delta(|\mathbf{r}|-a)}{4\pi}\left[ 1-\omega-\frac{\mathbf{r}\cdot\mathbf{W}}{a}+\frac{r_ir_j\partial_jW_i}{2a}+\frac{(\mathbf{r}\cdot\mathbf{W})^2}{2a^2} \right]\times \\ \nonumber
& &\left[ q(\mathbf{x},n)-r_k\partial_kq(\mathbf{x},n)+\frac{r_kr_l}{2}\partial_k\partial_lq(\mathbf{x},n) \right] \\
&= &q(\mathbf{x},n) + \left[ -\omega+\frac{a^2}{6}\nabla^2+\frac{a}{6}\partial_iW_i+\frac{a}{3}W_i\partial_i+\frac{\mathbf{W}^2}{6} \right]q(\mathbf{x},n).
\label{eqn:microscopicDiffusionEqnApprox}
\end{eqnarray}
where we have used $\int \mathrm{d}^3\mathbf{r}~r_ir_j\delta(|\mathbf{r}|-a)=4\pi a^2\delta_{ij}/3$.

We now change from a discrete to a continuous chain parametrization converting arc-length coordinate $s = n/N$ by substituting $q(\mathbf{x},n+1)-q(\mathbf{x},n)\to[q(\mathbf{x},s+\delta s)-q(\mathbf{x},s)]/N\delta s$ and taking the limit $\delta s=1/N\to 0$:
\begin{equation}
\frac{\partial q(\mathbf{x},s)}{\partial s} = \frac{N}{6}(a\nabla+\mathbf{W})^2q(\mathbf{x},s)-wq(\mathbf{x},s)
\label{eqn:diffEqnAppendix}
\end{equation}
where $w({\bf x}) = N \omega({\bf x})$. Eqn. \ref{eqn:diffusionEqn2} is derived by the same procedure starting from 
\begin{equation}
q^\dagger(\mathbf{x},n-1) = \int\mathrm{d}^3\mathbf{x}~p_p(\mathbf{x}+\mathbf{r},\mathbf{x})q^\dagger(\mathbf{x}+\mathbf{r},n)
\label{eqn:microscopicDiffusionEqn2}
\end{equation}.

We can also follow this procedure to derive the diffusion equation with nematic orientational interactions, with mean-field orientation interactions described by the self-consistent tensor $W_{ij}({\bf x})$. In this case, the transfer probability is
\begin{equation}
p_n(\mathbf{x}',\mathbf{x}) = \frac{\delta(|\mathbf{x}-\mathbf{x}'|-a)}{4\pi a^2}\exp{\left[-w(\bar{\mathbf{x}})-W_{ij}(\bar{\mathbf{x}})\left(\frac{(x_i-x_i')(x_j-x_j')}{a^2}-\frac{\delta_{ij}}{3}\right)\right]}
\label{eqn:transferProbabilityNematic}
\end{equation}
which we expand as 
\begin{equation}
p_n(\mathbf{x}-\mathbf{r},\mathbf{x}) \approx \left[ 1-w-\left( W_{ij}-\frac{r_k}{2}\partial_kW_{ij} \right)\left( \frac{r_ir_j}{a^2}-\frac{\delta_{ij}}{3} \right) \right]\delta(|{\bf r}| - a)
\label{eqn:transferProbabilityNematicExpansion}
\end{equation}
and continue as above to arrive at the effective diffusion equation for nematic segments, using the additional identity $\int \mathrm{d}^3\mathbf{r}~r_ir_jr_kr_l\delta(|\mathbf{r}|-a)=4\pi a^4(\delta_{ij}\delta_{kl}+\delta_{ik}\delta_{jl}+\delta_{il}\delta_{jk})/15$.  The modified diffusion equation can be derived from the freely joined chain model, resulting in
\begin{equation}
\frac{\partial q}{\partial s}=\begin{cases}
    \frac{Na^2}{6}\nabla^2q-\frac{Na^2}{15}\partial_i\tilde{W}_{ij}\partial_jq-w_\mathrm{A}(\mathbf{x})q & \text{if $s<f$}\\
    \frac{Na^2}{6}\nabla^2q-w_\mathrm{B}(\mathbf{x})q & \text{if $s>f$}
  \end{cases}
\label{eqn:diffusionEqnTensor}
\end{equation}
where $\tilde{W}_{ij} = 1/2(W_{ij}+W_{ji})-\delta_{ij}W_{kk}/3$ is the traceless symmetric part of $W_{ij}$, and similarly for $q^\dagger$, with $q\to q^\dagger$ and $\partial/\partial s \to -\partial/\partial s$.

The expressions for the polar and nematic order parameters also derive from microscopic chain statistics. The polar order parameter $\mathbf{t}(\mathbf{x})$ is given by integrating the probability that a segment will extend from $\mathbf{x}-\mathbf{r}/2$ to $\mathbf{x}+\mathbf{r}/2$ multiplied by its average orientation:
\begin{eqnarray}
\nonumber
\mathbf{t}^\alpha(\mathbf{x}) &=& \frac{n}{\rho_0 \mathcal{Q}}~\sum_{n \in \alpha}\int\mathrm{d}^3\mathbf{r}~q(\mathbf{x}-\mathbf{r}/2,n-1)q^\dagger(\mathbf{x}+\mathbf{r}/2,n+1)p_p(\mathbf{x}-\mathbf{r}/2,\mathbf{x}+\mathbf{r}/2)\frac{\mathbf{r}}{a} \\
&=& \frac{V}{6\mathcal{Q}}\int_\alpha\mathrm{d}s~\Big(q \nabla q^\dagger - q^\dagger \nabla q - 2 {\bf W}({\bf x})q^\dagger q \Big) 
\label{eqn:tFromStatistics}
\end{eqnarray}
where $\alpha$ identifies a monomer species, and the second line follows from the expansion of chain distributions to first order in ${\bf r}/2$ and averaging over orientation. The nematic order parameter is derives similarly:
\begin{eqnarray}
\nonumber
  Q^\alpha_{ij} ({\bf x}) &=&  \frac{n}{\rho_0  \mathcal{Q}}~\sum_{n \in \alpha} \int\mathrm{d}^3\mathbf{r}~q(\mathbf{x}-\mathbf{r}/2,n-1)q^\dagger(\mathbf{x}+\mathbf{r}/2,n+1)p_n(\mathbf{x}-\mathbf{r}/2,\mathbf{x}+\mathbf{r}/2)\Big(\frac{r_i r_j}{a^2}-\frac{\delta_{ij}}{3} \Big) \\ \nonumber
  &=& \frac{V}{60\mathcal{Q}}\int_\alpha\mathrm{d}s~\bigg[ q\partial_i\partial_jq^\dagger + q^\dagger\partial_i\partial_jq -\partial_iq\partial_jq^\dagger-\partial_iq^\dagger\partial_jq \\
 && \ \ \ \ \ \ \ \ \ \ \ \ \ \ \ \  -\frac{\delta_{ij}}{3}\left(q\nabla^2q^\dagger+q^\dagger\nabla^2q-2\nabla q\cdot\nabla q^\dagger\right)-8\tilde{W}_{ij}qq^\dagger \bigg] .
   \label{eqn:nematicOP}
\end{eqnarray}
an expression which is manifestly symmetric and traceless.

\section{Rescaled SCF equations}\label{appendix:rescaledSummary}

Applying the rescalings defined in section \ref{sec:scfTheory} gives the following set of rescaled polar SCF equations, which are all $N$-independent:
\begin{align}
\frac{\partial q}{\partial s}=&\begin{cases}
    \frac{1}{6}\left(\bar{\nabla}+\bar{\mathbf{W}}(\bar{\mathbf{x}})\right)^2q-w_\mathrm{A}(\bar{\mathbf{x}})q & \text{if $s<f$}\\
    \frac{1}{6}\bar{\nabla}^2q-w_\mathrm{B}(\bar{\mathbf{x}})q & \text{if $s>f$}
  \end{cases} \\
-\frac{\partial q^\dagger}{\partial s}=&\begin{cases}
    \frac{1}{6}\left(\bar{\nabla}-\bar{\mathbf{W}}(\bar{\mathbf{x}})\right)^2q^\dagger-w_\mathrm{A}(\bar{\mathbf{x}})q^\dagger & \text{if $s<f$}\\
    \frac{1}{6}\bar{\nabla}^2q^\dagger-w_\mathrm{B}(\bar{\mathbf{x}})q^\dagger & \text{if $s>f$}
  \end{cases} \\
\bar{\mathcal{Q}}=&\int\mathrm{d}^3\bar{\mathbf{x}}~q(\bar{\mathbf{x}},s)q^\dagger(\bar{\mathbf{x}},s) \\
\phi_\mathrm{A}(\bar{\mathbf{x}})=&\frac{\bar{V}}{\bar{\mathcal{Q}}}\int_0^f \mathrm{d}s~q(\bar{\mathbf{x}},s)q^\dagger(\bar{\mathbf{x}},s) \\
\phi_\mathrm{B}(\bar{\mathbf{x}})=&\frac{\bar{V}}{\bar{\mathcal{Q}}}\int_f^1 \mathrm{d}s~q(\bar{\mathbf{x}},s)q^\dagger(\bar{\mathbf{x}},s) \\
\bar{\mathbf{t}}(\bar{\mathbf{x}})=&\frac{\bar{V}}{6\bar{\mathcal{Q}}}\int_0^f \mathrm{d}s~q\bar{\nabla} q^\dagger -q^\dagger\bar{\nabla} q -2\bar{\mathbf{W}}qq^\dagger \\
w_{\mathrm{A}}(\bar{\mathbf{x}})=&\bar{\chi}\phi_{\mathrm{B}}(\bar{\mathbf{x}})+\xi(\bar{\mathbf{x}}) \\
w_{\mathrm{B}}(\bar{\mathbf{x}})=&\bar{\chi}\phi_{\mathrm{A}}(\bar{\mathbf{x}})+\xi(\bar{\mathbf{x}}) \\
   \bar{\mathbf{W}}(\bar{\mathbf{x}})=&\bar{A}\bar{\mathbf{t}}(\bar{\mathbf{x}})+\bar{C}|\bar{\mathbf{t}}(\bar{\mathbf{x}})|^2\bar{\mathbf{t}}(\bar{\mathbf{x}})-\bar{K}_1\bar{\nabla}(\bar{\nabla}\cdot\bar{\mathbf{t}})-\bar{K}_2\bar{\nabla}\times(\bar{\nabla}\times\bar{\mathbf{t}})+2\bar{K}_2\bar{q}_0\bar{\nabla}\times\bar{\mathbf{t}}.
\end{align}

Applying the rescalings to the nematic equations does not remove all $N$-dependence. The free energy and diffusion equations remain explicitly $N$-dependent:
\begin{align}
\label{eq:dqdsQ}
\frac{\partial q}{\partial s}=&\begin{cases}
    \frac{1}{6}\bar{\nabla}^2q-\frac{1}{15N}\bar{\partial}_i\bar{W}_{ij}\bar{\partial}_jq-w_\mathrm{A}(\mathbf{x})q & \text{if $s<f$}\\
    \frac{1}{6}\bar{\nabla}^2q-w_\mathrm{B}(\mathbf{x})q & \text{if $s>f$}
  \end{cases} \\
   F_{\mathrm{n}}^\mathrm{local}=&\frac{1}{N\bar{V}}\int\mathrm{d}^3\bar{\mathbf{x}}~\left[\bar{A}\mathrm{tr}(\bar{Q}^2)+\frac{2\bar{B}}{3}\mathrm{tr}(\bar{Q}^3)+\bar{C}\mathrm{tr}(\bar{Q}^4)\right] \\
   F_{\mathrm{n}}^\mathrm{grad}=&\frac{1}{N\bar{V}}\int\mathrm{d}^3\mathbf{x}\left[\bar{K}_1(\bar{\partial}_i \bar{Q}_{ij})^2+\bar{K}_2(\epsilon_{ikl}\bar{\partial}_k \bar{Q}_{lj})^2+2\bar{K}_2\bar{q}_0\bar{Q}_{ij}\epsilon_{ikl}\bar{\partial}_k \bar{Q}_{lj}\right] \\
   \label{eq:SQ}
   S=&\ln{\bar{\mathcal{Q}}}+\frac{1}{V}\int\mathrm{d}^3\bar{\mathbf{x}}~\left(w_\mathrm{A}\phi_\mathrm{A}+w_\mathrm{B}\phi_\mathrm{B}+\frac{1}{N}\bar{W}_{ij}\bar{Q}_{ij}\right).
\end{align}
The remaining equations are $N$-independent:
\begin{align}
\label{eq:Q}
\bar{Q}_{ij} =& \frac{\bar{V}}{60\bar{\mathcal{Q}}}\int\mathrm{d}s~\bigg[ q\bar{\partial}_i\bar{\partial}_jq^\dagger + q^\dagger\bar{\partial}_i\bar{\partial}_jq -\bar{\partial}_iq\bar{\partial}_jq^\dagger-\bar{\partial}_iq^\dagger\bar{\partial}_jq \\
   &-\frac{\delta_{ij}}{3}\left(q\bar{\nabla}^2q^\dagger+q^\dagger\bar{\nabla}^2q-2\bar{\nabla} q\cdot\bar{\nabla} q^\dagger\right)-8\tilde{\bar{W}}_{ij}qq^\dagger \bigg] \\
\bar{W}_{ij} =& 2\bar{A}\bar{Q}_{ij}+2\bar{B}\bar{Q}_{ik}\bar{Q}_{kj}+4\bar{C}\bar{Q}_{ik}\bar{Q}_{kl}\bar{Q}_{lj}+(\bar{K}_2-\bar{K}_1)\bar{\partial}_i \bar{\partial}_k \bar{Q}_{kj}-\bar{K}_2\bar{\partial}_k^2\bar{Q}_{ij}+2\bar{K}_2\bar{q}_0\epsilon_{ikl}\bar{\partial}_k \bar{Q}_{lj}.
\end{align}
All other equations are the same as in the polar theory.

%
%
%

%
%

\bibliography{bcpBulkPaper}

\begin{thebibliography}{36}%
\makeatletter
\providecommand \@ifxundefined [1]{%
 \@ifx{#1\undefined}
}%
\providecommand \@ifnum [1]{%
 \ifnum #1\expandafter \@firstoftwo
 \else \expandafter \@secondoftwo
 \fi
}%
\providecommand \@ifx [1]{%
 \ifx #1\expandafter \@firstoftwo
 \else \expandafter \@secondoftwo
 \fi
}%
\providecommand \natexlab [1]{#1}%
\providecommand \enquote  [1]{``#1''}%
\providecommand \bibnamefont  [1]{#1}%
\providecommand \bibfnamefont [1]{#1}%
\providecommand \citenamefont [1]{#1}%
\providecommand \href@noop [0]{\@secondoftwo}%
\providecommand \href [0]{\begingroup \@sanitize@url \@href}%
\providecommand \@href[1]{\@@startlink{#1}\@@href}%
\providecommand \@@href[1]{\endgroup#1\@@endlink}%
\providecommand \@sanitize@url [0]{\catcode `\\12\catcode `\$12\catcode
  `\&12\catcode `\#12\catcode `\^12\catcode `\_12\catcode `\%12\relax}%
\providecommand \@@startlink[1]{}%
\providecommand \@@endlink[0]{}%
\providecommand \url  [0]{\begingroup\@sanitize@url \@url }%
\providecommand \@url [1]{\endgroup\@href {#1}{\urlprefix }}%
\providecommand \urlprefix  [0]{URL }%
\providecommand \Eprint [0]{\href }%
\providecommand \doibase [0]{http://dx.doi.org/}%
\providecommand \selectlanguage [0]{\@gobble}%
\providecommand \bibinfo  [0]{\@secondoftwo}%
\providecommand \bibfield  [0]{\@secondoftwo}%
\providecommand \translation [1]{[#1]}%
\providecommand \BibitemOpen [0]{}%
\providecommand \bibitemStop [0]{}%
\providecommand \bibitemNoStop [0]{.\EOS\space}%
\providecommand \EOS [0]{\spacefactor3000\relax}%
\providecommand \BibitemShut  [1]{\csname bibitem#1\endcsname}%
\let\auto@bib@innerbib\@empty
\bibitem [{\citenamefont {Bates}\ and\ \citenamefont
  {Fredrickson}(1990)}]{Bates1990}%
  \BibitemOpen
  \bibfield  {author} {\bibinfo {author} {\bibfnamefont {F.~S.}\ \bibnamefont
  {Bates}}\ and\ \bibinfo {author} {\bibfnamefont {G.~H.}\ \bibnamefont
  {Fredrickson}},\ }\bibfield  {title} {\enquote {\bibinfo {title} {{Block
  Copolymer Thermodynamics: Theory and Experiment}},}\ }\href {\doibase
  10.1146/annurev.pc.41.100190.002521} {\bibfield  {journal} {\bibinfo
  {journal} {Annual Review of Physical Chemistry}\ }\textbf {\bibinfo {volume}
  {41}},\ \bibinfo {pages} {525--557} (\bibinfo {year} {1990})}\BibitemShut
  {NoStop}%
\bibitem [{\citenamefont {Grason}(2006)}]{Grason2006}%
  \BibitemOpen
  \bibfield  {author} {\bibinfo {author} {\bibfnamefont {G.~M.}\ \bibnamefont
  {Grason}},\ }\bibfield  {title} {\enquote {\bibinfo {title} {{The packing of
  soft materials: Molecular asymmetry, geometric frustration and optimal
  lattices in block copolymer melts}},}\ }\href {\doibase
  10.1016/j.physrep.2006.08.001} {\bibfield  {journal} {\bibinfo  {journal}
  {Physics Reports}\ }\textbf {\bibinfo {volume} {433}},\ \bibinfo {pages}
  {1--64} (\bibinfo {year} {2006})}\BibitemShut {NoStop}%
\bibitem [{\citenamefont {Matsen}(2012)}]{Matsen2012}%
  \BibitemOpen
  \bibfield  {author} {\bibinfo {author} {\bibfnamefont {M.~W.}\ \bibnamefont
  {Matsen}},\ }\bibfield  {title} {\enquote {\bibinfo {title} {{Effect of
  Architecture on the Phase Behavior of AB-Type Block Copolymer Melts}},}\
  }\href {\doibase 10.1021/ma202782s} {\bibfield  {journal} {\bibinfo
  {journal} {Macromolecules}\ }\textbf {\bibinfo {volume} {45}},\ \bibinfo
  {pages} {2161--2165} (\bibinfo {year} {2012})}\BibitemShut {NoStop}%
\bibitem [{\citenamefont {Leibler}(1980)}]{Leibler1980}%
  \BibitemOpen
  \bibfield  {author} {\bibinfo {author} {\bibfnamefont {L.}~\bibnamefont
  {Leibler}},\ }\bibfield  {title} {\enquote {\bibinfo {title} {{Theory of
  Microphase Separation in Block Copolymers}},}\ }\href {\doibase
  10.1021/ma60078a047} {\bibfield  {journal} {\bibinfo  {journal}
  {Macromolecules}\ }\textbf {\bibinfo {volume} {13}},\ \bibinfo {pages}
  {1602--1617} (\bibinfo {year} {1980})},\ \Eprint
  {http://arxiv.org/abs/0402594v3} {arXiv:0402594v3 [arXiv:cond-mat]}
  \BibitemShut {NoStop}%
\bibitem [{\citenamefont {Helfand}(1975)}]{Helfand1975}%
  \BibitemOpen
  \bibfield  {author} {\bibinfo {author} {\bibfnamefont {E.}~\bibnamefont
  {Helfand}},\ }\bibfield  {title} {\enquote {\bibinfo {title} {{Theory of
  inhomogeneous polymers: Fundamentals of the Gaussian random-walk model}},}\
  }\href {\doibase 10.1063/1.430517} {\bibfield  {journal} {\bibinfo  {journal}
  {The Journal of Chemical Physics}\ }\textbf {\bibinfo {volume} {62}},\
  \bibinfo {pages} {999} (\bibinfo {year} {1975})}\BibitemShut {NoStop}%
\bibitem [{\citenamefont {Matsen}(2002)}]{Matsen2002}%
  \BibitemOpen
  \bibfield  {author} {\bibinfo {author} {\bibfnamefont {M.~W.}\ \bibnamefont
  {Matsen}},\ }\bibfield  {title} {\enquote {\bibinfo {title} {{The standard
  Gaussian model for block copolymer melts}},}\ }\href {\doibase
  10.1088/0953-8984/14/2/201} {\bibfield  {journal} {\bibinfo  {journal} {J.
  Phys.: Condens. Matter}\ }\textbf {\bibinfo {volume} {14}},\ \bibinfo {pages}
  {21--47} (\bibinfo {year} {2002})}\BibitemShut {NoStop}%
\bibitem [{\citenamefont {Khandpur}\ \emph {et~al.}(1995)\citenamefont
  {Khandpur}, \citenamefont {F{\"{o}}rster}, \citenamefont {Bates},
  \citenamefont {Hamley}, \citenamefont {Ryan}, \citenamefont {Bras},
  \citenamefont {Almdal},\ and\ \citenamefont {Mortensen}}]{Khandpur1995}%
  \BibitemOpen
  \bibfield  {author} {\bibinfo {author} {\bibfnamefont {A.~K.}\ \bibnamefont
  {Khandpur}}, \bibinfo {author} {\bibfnamefont {S.}~\bibnamefont
  {F{\"{o}}rster}}, \bibinfo {author} {\bibfnamefont {F.~S.}\ \bibnamefont
  {Bates}}, \bibinfo {author} {\bibfnamefont {I.~W.}\ \bibnamefont {Hamley}},
  \bibinfo {author} {\bibfnamefont {A.~J.}\ \bibnamefont {Ryan}}, \bibinfo
  {author} {\bibfnamefont {W.}~\bibnamefont {Bras}}, \bibinfo {author}
  {\bibfnamefont {K.}~\bibnamefont {Almdal}}, \ and\ \bibinfo {author}
  {\bibfnamefont {K.}~\bibnamefont {Mortensen}},\ }\bibfield  {title} {\enquote
  {\bibinfo {title} {{Polyisoprene-Polystyrene Diblock Copolymer Phase Diagram
  near the Order-Disorder Transition}},}\ }\href {\doibase 10.1021/ma00130a012}
  {\bibfield  {journal} {\bibinfo  {journal} {Macromolecules}\ }\textbf
  {\bibinfo {volume} {28}},\ \bibinfo {pages} {8796--8806} (\bibinfo {year}
  {1995})}\BibitemShut {NoStop}%
\bibitem [{\citenamefont {Fredrickson}\ and\ \citenamefont
  {Helfand}(1987)}]{Fredrickson1987}%
  \BibitemOpen
  \bibfield  {author} {\bibinfo {author} {\bibfnamefont {G.~H.}\ \bibnamefont
  {Fredrickson}}\ and\ \bibinfo {author} {\bibfnamefont {E.}~\bibnamefont
  {Helfand}},\ }\bibfield  {title} {\enquote {\bibinfo {title} {{Fluctuation
  effects in the theory of microphase separation in block copolymers}},}\
  }\href {\doibase 10.1063/1.453566} {\bibfield  {journal} {\bibinfo  {journal}
  {The Journal of Chemical Physics}\ }\textbf {\bibinfo {volume} {87}},\
  \bibinfo {pages} {697--705} (\bibinfo {year} {1987})}\BibitemShut {NoStop}%
\bibitem [{\citenamefont {Matsen}\ and\ \citenamefont
  {Schick}(1994)}]{Matsen1994}%
  \BibitemOpen
  \bibfield  {author} {\bibinfo {author} {\bibfnamefont {M.~W.}\ \bibnamefont
  {Matsen}}\ and\ \bibinfo {author} {\bibfnamefont {M.}~\bibnamefont
  {Schick}},\ }\bibfield  {title} {\enquote {\bibinfo {title} {{Stable and
  unstable phases of a diblock copolymer melt}},}\ }\href {\doibase
  10.1103/PhysRevLett.72.2660} {\bibfield  {journal} {\bibinfo  {journal}
  {Physical Review Letters}\ }\textbf {\bibinfo {volume} {72}},\ \bibinfo
  {pages} {2660--2663} (\bibinfo {year} {1994})}\BibitemShut {NoStop}%
\bibitem [{\citenamefont {Milner}(1994)}]{Milner1994}%
  \BibitemOpen
  \bibfield  {author} {\bibinfo {author} {\bibfnamefont {S.~T.}\ \bibnamefont
  {Milner}},\ }\bibfield  {title} {\enquote {\bibinfo {title} {{Chain
  Architecture and Asymmetry in Copolymer Microphases}},}\ }\href {\doibase
  10.1021/ma00086a057} {\bibfield  {journal} {\bibinfo  {journal}
  {Macromolecules}\ }\textbf {\bibinfo {volume} {27}},\ \bibinfo {pages}
  {2333--2335} (\bibinfo {year} {1994})}\BibitemShut {NoStop}%
\bibitem [{\citenamefont {Burger}, \citenamefont {Ruland},\ and\ \citenamefont
  {Semenov}(1990)}]{Burger1990}%
  \BibitemOpen
  \bibfield  {author} {\bibinfo {author} {\bibfnamefont {C.}~\bibnamefont
  {Burger}}, \bibinfo {author} {\bibfnamefont {W.}~\bibnamefont {Ruland}}, \
  and\ \bibinfo {author} {\bibfnamefont {A.~N.}\ \bibnamefont {Semenov}},\
  }\bibfield  {title} {\enquote {\bibinfo {title} {{Polydispersity effects on
  the microphase-separation transition in block copolymers}},}\ }\href
  {\doibase 10.1021/ma00215a021} {\bibfield  {journal} {\bibinfo  {journal}
  {Macromolecules}\ }\textbf {\bibinfo {volume} {23}},\ \bibinfo {pages}
  {3339--3346} (\bibinfo {year} {1990})}\BibitemShut {NoStop}%
\bibitem [{\citenamefont {Chen}(2016)}]{Chen2016}%
  \BibitemOpen
  \bibfield  {author} {\bibinfo {author} {\bibfnamefont {J.~Z.~Y.}\
  \bibnamefont {Chen}},\ }\bibfield  {title} {\enquote {\bibinfo {title}
  {{Theory of wormlike polymer chains in confinement}},}\ }\href {\doibase
  10.1016/j.progpolymsci.2015.09.002} {\bibfield  {journal} {\bibinfo
  {journal} {Progress in Polymer Science}\ }\textbf {\bibinfo {volume}
  {54-55}},\ \bibinfo {pages} {3--46} (\bibinfo {year} {2016})}\BibitemShut
  {NoStop}%
\bibitem [{\citenamefont {Pryamitsyn}\ and\ \citenamefont
  {Ganesan}(2004)}]{Pryamitsyn2004}%
  \BibitemOpen
  \bibfield  {author} {\bibinfo {author} {\bibfnamefont {V.}~\bibnamefont
  {Pryamitsyn}}\ and\ \bibinfo {author} {\bibfnamefont {V.}~\bibnamefont
  {Ganesan}},\ }\bibfield  {title} {\enquote {\bibinfo {title} {{Self-assembly
  of rod--coil block copolymers}},}\ }\href {\doibase 10.1063/1.1649729}
  {\bibfield  {journal} {\bibinfo  {journal} {The Journal of Chemical Physics}\
  }\textbf {\bibinfo {volume} {120}},\ \bibinfo {pages} {5824--5838} (\bibinfo
  {year} {2004})}\BibitemShut {NoStop}%
\bibitem [{\citenamefont {Matsen}\ and\ \citenamefont
  {Barrett}(1998)}]{Matsen1998}%
  \BibitemOpen
  \bibfield  {author} {\bibinfo {author} {\bibfnamefont {M.~W.}\ \bibnamefont
  {Matsen}}\ and\ \bibinfo {author} {\bibfnamefont {C.}~\bibnamefont
  {Barrett}},\ }\bibfield  {title} {\enquote {\bibinfo {title}
  {{Liquid-crystalline behavior of rod-coil diblock copolymers}},}\ }\href
  {\doibase 10.1063/1.477011} {\bibfield  {journal} {\bibinfo  {journal}
  {Journal of Chemical Physics}\ }\textbf {\bibinfo {volume} {109}},\ \bibinfo
  {pages} {4108--4118} (\bibinfo {year} {1998})}\BibitemShut {NoStop}%
\bibitem [{\citenamefont {Matsen}(1996)}]{Matsen1996}%
  \BibitemOpen
  \bibfield  {author} {\bibinfo {author} {\bibfnamefont {M.~W.}\ \bibnamefont
  {Matsen}},\ }\bibfield  {title} {\enquote {\bibinfo {title} {{Melts of
  semiflexible diblock copolymer}},}\ }\href {\doibase Doi 10.1063/1.471481}
  {\bibfield  {journal} {\bibinfo  {journal} {J. Chem. Phys.}\ }\textbf
  {\bibinfo {volume} {104}},\ \bibinfo {pages} {7758--7764} (\bibinfo {year}
  {1996})}\BibitemShut {NoStop}%
\bibitem [{\citenamefont {Jiang}, \citenamefont {Zhang},\ and\ \citenamefont
  {Chen}(2011)}]{Jiang2011}%
  \BibitemOpen
  \bibfield  {author} {\bibinfo {author} {\bibfnamefont {Y.}~\bibnamefont
  {Jiang}}, \bibinfo {author} {\bibfnamefont {W.~Y.}\ \bibnamefont {Zhang}}, \
  and\ \bibinfo {author} {\bibfnamefont {J.~Z.}\ \bibnamefont {Chen}},\
  }\bibfield  {title} {\enquote {\bibinfo {title} {{Dependence of the
  disorder-lamellar stability boundary of a melt of asymmetric wormlike AB
  diblock copolymers on the chain rigidity}},}\ }\href {\doibase
  10.1103/PhysRevE.84.041803} {\bibfield  {journal} {\bibinfo  {journal}
  {Physical Review E - Statistical, Nonlinear, and Soft Matter Physics}\
  }\textbf {\bibinfo {volume} {84}},\ \bibinfo {pages} {1--10} (\bibinfo {year}
  {2011})}\BibitemShut {NoStop}%
\bibitem [{\citenamefont {Jiang}\ and\ \citenamefont {Chen}(2013)}]{Jiang2013}%
  \BibitemOpen
  \bibfield  {author} {\bibinfo {author} {\bibfnamefont {Y.}~\bibnamefont
  {Jiang}}\ and\ \bibinfo {author} {\bibfnamefont {J.~Z.~Y.}\ \bibnamefont
  {Chen}},\ }\bibfield  {title} {\enquote {\bibinfo {title} {{Influence of
  chain rigidity on the phase behavior of wormlike diblock copolymers}},}\
  }\href {\doibase 10.1103/PhysRevLett.110.138305} {\bibfield  {journal}
  {\bibinfo  {journal} {Physical Review Letters}\ }\textbf {\bibinfo {volume}
  {110}},\ \bibinfo {pages} {1--5} (\bibinfo {year} {2013})}\BibitemShut
  {NoStop}%
\bibitem [{\citenamefont {Singh}\ \emph {et~al.}(1994)\citenamefont {Singh},
  \citenamefont {Goulian}, \citenamefont {Liu},\ and\ \citenamefont
  {Fredrickson}}]{Singh1994}%
  \BibitemOpen
  \bibfield  {author} {\bibinfo {author} {\bibfnamefont {C.}~\bibnamefont
  {Singh}}, \bibinfo {author} {\bibfnamefont {M.}~\bibnamefont {Goulian}},
  \bibinfo {author} {\bibfnamefont {A.~J.}\ \bibnamefont {Liu}}, \ and\
  \bibinfo {author} {\bibfnamefont {G.~H.}\ \bibnamefont {Fredrickson}},\
  }\bibfield  {title} {\enquote {\bibinfo {title} {{Phase-Behavior of
  Semiflexible Diblock Copolymers}},}\ }\href {\doibase 10.1021/ma00089a013}
  {\bibfield  {journal} {\bibinfo  {journal} {Macromolecules}\ }\textbf
  {\bibinfo {volume} {27}},\ \bibinfo {pages} {2974--2986} (\bibinfo {year}
  {1994})}\BibitemShut {NoStop}%
\bibitem [{\citenamefont {Netz}\ and\ \citenamefont {Schick}(1996)}]{Netz1996}%
  \BibitemOpen
  \bibfield  {author} {\bibinfo {author} {\bibfnamefont {R.}~\bibnamefont
  {Netz}}\ and\ \bibinfo {author} {\bibfnamefont {M.}~\bibnamefont {Schick}},\
  }\bibfield  {title} {\enquote {\bibinfo {title} {{Liquid-Crystalline Phases
  of Semiflexible Diblock Copolymer Melts}},}\ }\href {\doibase
  10.1103/PhysRevLett.77.302} {\bibfield  {journal} {\bibinfo  {journal}
  {Physical Review Letters}\ }\textbf {\bibinfo {volume} {77}},\ \bibinfo
  {pages} {302--305} (\bibinfo {year} {1996})}\BibitemShut {NoStop}%
\bibitem [{\citenamefont {Greco}\ \emph {et~al.}(2016)\citenamefont {Greco},
  \citenamefont {Jiang}, \citenamefont {Chen}, \citenamefont {Kremer},\ and\
  \citenamefont {Daoulas}}]{Greco2016}%
  \BibitemOpen
  \bibfield  {author} {\bibinfo {author} {\bibfnamefont {C.}~\bibnamefont
  {Greco}}, \bibinfo {author} {\bibfnamefont {Y.}~\bibnamefont {Jiang}},
  \bibinfo {author} {\bibfnamefont {J.~Z.~Y.}\ \bibnamefont {Chen}}, \bibinfo
  {author} {\bibfnamefont {K.}~\bibnamefont {Kremer}}, \ and\ \bibinfo {author}
  {\bibfnamefont {K.~C.}\ \bibnamefont {Daoulas}},\ }\bibfield  {title}
  {\enquote {\bibinfo {title} {{Maier-Saupe model of polymer nematics:
  Comparing free energies calculated with Self Consistent Field theory and
  Monte Carlo simulations}},}\ }\href {\doibase 10.1063/1.4966919} {\bibfield
  {journal} {\bibinfo  {journal} {The Journal of Chemical Physics}\ }\textbf
  {\bibinfo {volume} {145}},\ \bibinfo {pages} {184901} (\bibinfo {year}
  {2016})}\BibitemShut {NoStop}%
\bibitem [{\citenamefont {Jiang}, \citenamefont {Li},\ and\ \citenamefont
  {Chen}(2016)}]{Jiang2016}%
  \BibitemOpen
  \bibfield  {author} {\bibinfo {author} {\bibfnamefont {Y.}~\bibnamefont
  {Jiang}}, \bibinfo {author} {\bibfnamefont {S.}~\bibnamefont {Li}}, \ and\
  \bibinfo {author} {\bibfnamefont {J.~Z.}\ \bibnamefont {Chen}},\ }\bibfield
  {title} {\enquote {\bibinfo {title} {{Perspective: parameters in a
  self-consistent field theory of multicomponent wormlike-copolymer melts}},}\
  }\href {\doibase 10.1140/epje/i2016-16091-8} {\bibfield  {journal} {\bibinfo
  {journal} {The European Physical Journal E}\ }\textbf {\bibinfo {volume}
  {39}},\ \bibinfo {pages} {91} (\bibinfo {year} {2016})}\BibitemShut {NoStop}%
\bibitem [{\citenamefont {Prasad}\ \emph {et~al.}(2017)\citenamefont {Prasad},
  \citenamefont {Seo}, \citenamefont {Hall},\ and\ \citenamefont
  {Grason}}]{Prasad2017}%
  \BibitemOpen
  \bibfield  {author} {\bibinfo {author} {\bibfnamefont {I.}~\bibnamefont
  {Prasad}}, \bibinfo {author} {\bibfnamefont {Y.}~\bibnamefont {Seo}},
  \bibinfo {author} {\bibfnamefont {L.~M.}\ \bibnamefont {Hall}}, \ and\
  \bibinfo {author} {\bibfnamefont {G.~M.}\ \bibnamefont {Grason}},\ }\bibfield
   {title} {\enquote {\bibinfo {title} {{Intradomain Textures in Block
  Copolymers: Multizone Alignment and Biaxiality}},}\ }\href {\doibase
  10.1103/PhysRevLett.118.247801} {\bibfield  {journal} {\bibinfo  {journal}
  {Physical Review Letters}\ }\textbf {\bibinfo {volume} {118}},\ \bibinfo
  {pages} {247801} (\bibinfo {year} {2017})},\ \Eprint
  {http://arxiv.org/abs/1612.07994} {arXiv:1612.07994} \BibitemShut {NoStop}%
\bibitem [{\citenamefont {Carton}\ and\ \citenamefont
  {Leibler}(1990)}]{Carton1990}%
  \BibitemOpen
  \bibfield  {author} {\bibinfo {author} {\bibfnamefont {J.-P.}\ \bibnamefont
  {Carton}}\ and\ \bibinfo {author} {\bibfnamefont {L.}~\bibnamefont
  {Leibler}},\ }\bibfield  {title} {\enquote {\bibinfo {title}
  {Density-conformation coupling in macromolecular systems: polymer
  interfaces},}\ }\href {\doibase 10.1051/jphys:0199000510160168300} {\bibfield
   {journal} {\bibinfo  {journal} {Journal de Physique}\ }\textbf {\bibinfo
  {volume} {51}},\ \bibinfo {pages} {1683--1691} (\bibinfo {year}
  {1990})}\BibitemShut {NoStop}%
\bibitem [{\citenamefont {Szleifer}\ and\ \citenamefont
  {Widom}(1989)}]{Szleifer1989}%
  \BibitemOpen
  \bibfield  {author} {\bibinfo {author} {\bibfnamefont {I.}~\bibnamefont
  {Szleifer}}\ and\ \bibinfo {author} {\bibfnamefont {B.}~\bibnamefont
  {Widom}},\ }\bibfield  {title} {\enquote {\bibinfo {title} {{Structure and
  tension of the interface between dilute polymer solutions}},}\ }\href
  {\doibase 10.1063/1.456186} {\bibfield  {journal} {\bibinfo  {journal} {The
  Journal of Chemical Physics}\ }\textbf {\bibinfo {volume} {90}},\ \bibinfo
  {pages} {7524--7534} (\bibinfo {year} {1989})}\BibitemShut {NoStop}%
\bibitem [{\citenamefont {Morse}\ and\ \citenamefont
  {Fredrickson}(1994)}]{Morse1994}%
  \BibitemOpen
  \bibfield  {author} {\bibinfo {author} {\bibfnamefont {D.~C.}\ \bibnamefont
  {Morse}}\ and\ \bibinfo {author} {\bibfnamefont {G.~H.}\ \bibnamefont
  {Fredrickson}},\ }\bibfield  {title} {\enquote {\bibinfo {title}
  {{Semiflexible polymers near interfaces}},}\ }\href {\doibase
  10.1103/PhysRevLett.73.3235} {\bibfield  {journal} {\bibinfo  {journal}
  {Physical Review Letters}\ }\textbf {\bibinfo {volume} {73}},\ \bibinfo
  {pages} {3235--3238} (\bibinfo {year} {1994})}\BibitemShut {NoStop}%
\bibitem [{\citenamefont {Li}, \citenamefont {Jiang},\ and\ \citenamefont
  {Chen}(2016)}]{Li2016}%
  \BibitemOpen
  \bibfield  {author} {\bibinfo {author} {\bibfnamefont {S.}~\bibnamefont
  {Li}}, \bibinfo {author} {\bibfnamefont {Y.}~\bibnamefont {Jiang}}, \ and\
  \bibinfo {author} {\bibfnamefont {J.~Z.~Y.}\ \bibnamefont {Chen}},\
  }\bibfield  {title} {\enquote {\bibinfo {title} {{Complex liquid-crystal
  nanostructures in semiflexible ABC linear triblock copolymers: A
  self-consistent field theory}},}\ }\href {\doibase 10.1063/1.4967423}
  {\bibfield  {journal} {\bibinfo  {journal} {The Journal of Chemical Physics}\
  }\textbf {\bibinfo {volume} {145}},\ \bibinfo {pages} {184902} (\bibinfo
  {year} {2016})}\BibitemShut {NoStop}%
\bibitem [{\citenamefont {de~Gennes}\ and\ \citenamefont
  {Prost}(1993)}]{deGennes}%
  \BibitemOpen
  \bibfield  {author} {\bibinfo {author} {\bibfnamefont {P.~G.}\ \bibnamefont
  {de~Gennes}}\ and\ \bibinfo {author} {\bibfnamefont {J.}~\bibnamefont
  {Prost}},\ }\href@noop {} {\emph {\bibinfo {title} {The Physics of Liquid
  Crystals}}},\ \bibinfo {edition} {2nd}\ ed.\ (\bibinfo  {publisher}
  {Clarendon, Oxford},\ \bibinfo {year} {1993})\BibitemShut {NoStop}%
\bibitem [{\citenamefont {Zhao}, \citenamefont {Russell},\ and\ \citenamefont
  {Grason}(2012)}]{Zhao2012}%
  \BibitemOpen
  \bibfield  {author} {\bibinfo {author} {\bibfnamefont {W.}~\bibnamefont
  {Zhao}}, \bibinfo {author} {\bibfnamefont {T.~P.}\ \bibnamefont {Russell}}, \
  and\ \bibinfo {author} {\bibfnamefont {G.~M.}\ \bibnamefont {Grason}},\
  }\bibfield  {title} {\enquote {\bibinfo {title} {{Orientational interactions
  in block copolymer melts: self-consistent field theory.}}}\ }\href {\doibase
  10.1063/1.4752198} {\bibfield  {journal} {\bibinfo  {journal} {The Journal of
  chemical physics}\ }\textbf {\bibinfo {volume} {137}},\ \bibinfo {pages}
  {104911} (\bibinfo {year} {2012})}\BibitemShut {NoStop}%
\bibitem [{Note1()}]{Note1}%
  \BibitemOpen
  \bibinfo {note} {Note that the total derivative term proportional to $\DOTSI
  \intop \ilimits@ \protect \mathrm {d}^3\protect \mathbf {x} ~(\nabla \cdot
  {\protect \bf t})$ is also allowed by symmetry, but this vanishes for all
  periodic solutions of ${\protect \bf t}({\protect \bf x})$.}\BibitemShut
  {Stop}%
\bibitem [{\citenamefont {Zhao}, \citenamefont {Russell},\ and\ \citenamefont
  {Grason}(2013)}]{Zhao2013}%
  \BibitemOpen
  \bibfield  {author} {\bibinfo {author} {\bibfnamefont {W.}~\bibnamefont
  {Zhao}}, \bibinfo {author} {\bibfnamefont {T.~P.}\ \bibnamefont {Russell}}, \
  and\ \bibinfo {author} {\bibfnamefont {G.~M.}\ \bibnamefont {Grason}},\
  }\bibfield  {title} {\enquote {\bibinfo {title} {{Chirality in block
  copolymer melts: Mesoscopic helicity from intersegment twist}},}\ }\href
  {\doibase 10.1103/PhysRevLett.110.058301} {\bibfield  {journal} {\bibinfo
  {journal} {Physical Review Letters}\ }\textbf {\bibinfo {volume} {110}},\
  \bibinfo {pages} {1--5} (\bibinfo {year} {2013})},\ \Eprint
  {http://arxiv.org/abs/1210.1764} {arXiv:1210.1764} \BibitemShut {NoStop}%
\bibitem [{Note2()}]{Note2}%
  \BibitemOpen
  \bibinfo {note} {Such an effect can be seen considering the moments of 
  segment orientation, $\langle r \rangle$ and $\langle (\protect \mathaccentV 
  {hat}05E{r})_i (\protect \mathaccentV {hat}05E{r})_j \rangle$, for chain 
  held at fixed tension, with fixed contour length $Na$, but with variable 
  Kuhn length $a$.}\BibitemShut {NoStop}%
\bibitem [{\citenamefont {Rasmussen}\ and\ \citenamefont
  {Kalosakas}(2002)}]{Rasmussen2002}%
  \BibitemOpen
  \bibfield  {author} {\bibinfo {author} {\bibfnamefont {K.}~\bibnamefont
  {Rasmussen}}\ and\ \bibinfo {author} {\bibfnamefont {G.}~\bibnamefont
  {Kalosakas}},\ }\bibfield  {title} {\enquote {\bibinfo {title} {{Improved
  numerical algorithm for exploring block copolymer mesophases}},}\ }\href
  {\doibase 10.1002/polb.10238} {\bibfield  {journal} {\bibinfo  {journal}
  {Journal of Polymer Science, Part B: Polymer Physics}\ }\textbf {\bibinfo
  {volume} {40}},\ \bibinfo {pages} {1777--1783} (\bibinfo {year}
  {2002})}\BibitemShut {NoStop}%
\bibitem [{\citenamefont {Arora}\ \emph {et~al.}(2016)\citenamefont {Arora},
  \citenamefont {Qin}, \citenamefont {Morse}, \citenamefont {Delaney},
  \citenamefont {Fredrickson}, \citenamefont {Bates},\ and\ \citenamefont
  {Dorfman}}]{Aroraa}%
  \BibitemOpen
  \bibfield  {author} {\bibinfo {author} {\bibfnamefont {A.}~\bibnamefont
  {Arora}}, \bibinfo {author} {\bibfnamefont {J.}~\bibnamefont {Qin}}, \bibinfo
  {author} {\bibfnamefont {D.~C.}\ \bibnamefont {Morse}}, \bibinfo {author}
  {\bibfnamefont {K.~T.}\ \bibnamefont {Delaney}}, \bibinfo {author}
  {\bibfnamefont {G.~H.}\ \bibnamefont {Fredrickson}}, \bibinfo {author}
  {\bibfnamefont {F.~S.}\ \bibnamefont {Bates}}, \ and\ \bibinfo {author}
  {\bibfnamefont {K.~D.}\ \bibnamefont {Dorfman}},\ }\bibfield  {title}
  {\enquote {\bibinfo {title} {{Broadly Accessible Self-Consistent Field Theory
  for Block Polymer Materials Discovery}},}\ }\href {\doibase
  10.1021/acs.macromol.6b00107} {\bibfield  {journal} {\bibinfo  {journal}
  {Macromolecules}\ }\textbf {\bibinfo {volume} {49}},\ \bibinfo {pages}
  {4675--4690} (\bibinfo {year} {2016})}\BibitemShut {NoStop}%
\bibitem [{\citenamefont {Flannery}\ \emph {et~al.}(2007)\citenamefont
  {Flannery}, \citenamefont {Teukolsky}, \citenamefont {Press},\ and\
  \citenamefont {Vetterling}}]{numerical_recipes}%
  \BibitemOpen
  \bibfield  {author} {\bibinfo {author} {\bibfnamefont {B.~P.}\ \bibnamefont
  {Flannery}}, \bibinfo {author} {\bibfnamefont {S.}~\bibnamefont {Teukolsky}},
  \bibinfo {author} {\bibfnamefont {W.~H.}\ \bibnamefont {Press}}, \ and\
  \bibinfo {author} {\bibfnamefont {W.~T.}\ \bibnamefont {Vetterling}},\
  }\href@noop {} {\emph {\bibinfo {title} {Numerical Recipes in C: The Art of
  Scientific Computing}}}\ (\bibinfo  {publisher} {Cambridge University
  Press},\ \bibinfo {year} {2007})\BibitemShut {NoStop}%
\bibitem [{\citenamefont {Ding}\ and\ \citenamefont {Thomas}(1993)}]{Ding1993}%
  \BibitemOpen
  \bibfield  {author} {\bibinfo {author} {\bibfnamefont {D.~K.}\ \bibnamefont
  {Ding}}\ and\ \bibinfo {author} {\bibfnamefont {E.~L.}\ \bibnamefont
  {Thomas}},\ }\bibfield  {title} {\enquote {\bibinfo {title} {{Influence of
  elastic anisotropy on the structure of Neel inversion walls in liquid crystal
  polymers}},}\ }\href {\doibase 10.1021/ma00076a034} {\bibfield  {journal}
  {\bibinfo  {journal} {Macromolecules}\ }\textbf {\bibinfo {volume} {26}},\
  \bibinfo {pages} {6531--6535} (\bibinfo {year} {1993})}\BibitemShut {NoStop}%
\bibitem [{\citenamefont {Wang}\ \emph {et~al.}(2013)\citenamefont {Wang},
  \citenamefont {Kawakatsu}, \citenamefont {Chen},\ and\ \citenamefont
  {Lu}}]{Wang2013}%
  \BibitemOpen
  \bibfield  {author} {\bibinfo {author} {\bibfnamefont {S.-H.}\ \bibnamefont
  {Wang}}, \bibinfo {author} {\bibfnamefont {T.}~\bibnamefont {Kawakatsu}},
  \bibinfo {author} {\bibfnamefont {P.}~\bibnamefont {Chen}}, \ and\ \bibinfo
  {author} {\bibfnamefont {C.-Y.~D.}\ \bibnamefont {Lu}},\ }\bibfield  {title}
  {\enquote {\bibinfo {title} {{A density functional theory of chiral block
  copolymer melts}},}\ }\href {\doibase 10.1063/1.4802963} {\bibfield
  {journal} {\bibinfo  {journal} {The Journal of Chemical Physics}\ }\textbf
  {\bibinfo {volume} {138}},\ \bibinfo {pages} {194901} (\bibinfo {year}
  {2013})}\BibitemShut {NoStop}%
\end{thebibliography}%

\end{document}